\documentclass[a4paper,onecolumn,11pt,accepted=2024-06-24]{quantumarticle}
\pdfoutput=1
\usepackage[utf8]{inputenc}
\usepackage[english]{babel}
\usepackage[T1]{fontenc}
\usepackage{symbols}
\usepackage{xfrac}
\usepackage{soul}
\usepackage{amsfonts}
\usepackage{dsfont}
\usepackage{amsmath}
\usepackage{amssymb}
\usepackage{amsthm}
\usepackage{mathtools}
\usepackage{csquotes}
\usepackage[caption=false]{subfig}
\usepackage{bm}
\usepackage{hyperref}
\usepackage[
    style=phys,
    articletitle=false,
    biblabel=brackets,
    chaptertitle=false,
    pageranges=false,
    doi=false]{biblatex}

\newcommand{\new}[1]{{\color{black}#1}}
\newcommand{\old}[1]{}

\begin{document}

\title{Wigner Analysis of Particle Dynamics and Decoherence in Wide Nonharmonic Potentials}

\author{Andreu Riera-Campeny}
\affiliation{Institute for Quantum Optics and Quantum Information of the Austrian Academy of Sciences, A-6020 Innsbruck, Austria}
\affiliation{Institute for Theoretical Physics, University of Innsbruck, A-6020 Innsbruck, Austria}
\orcid{0000-0003-3260-993X}
\author{Marc Roda-Llordes}
\affiliation{Institute for Quantum Optics and Quantum Information of the Austrian Academy of Sciences, A-6020 Innsbruck, Austria}
\affiliation{Institute for Theoretical Physics, University of Innsbruck, A-6020 Innsbruck, Austria}
\orcid{0000-0002-7766-3190}
\author{Piotr T. Grochowski}
\affiliation{Institute for Quantum Optics and Quantum Information of the Austrian Academy of Sciences, A-6020 Innsbruck, Austria}
\affiliation{Institute for Theoretical Physics, University of Innsbruck, A-6020 Innsbruck, Austria}
\affiliation{Center for Theoretical Physics, Polish Academy of Sciences, Aleja Lotników 32/46, 02-668 Warsaw, Poland}
\orcid{0000-0002-9654-4824}
\author{Oriol Romero-Isart}
\affiliation{Institute for Quantum Optics and Quantum Information of the Austrian Academy of Sciences, A-6020 Innsbruck, Austria}
\affiliation{Institute for Theoretical Physics, University of Innsbruck, A-6020 Innsbruck, Austria}
\affiliation{ICFO - Institut de Ciències Fotòniques, The Barcelona Institute of Science and Technology, 08860 Castelldefels (Barcelona), Spain}
\affiliation{ICREA, Pg. Lluis Companys 23, 08010 Barcelona, Spain}
\orcid{0000-0003-4006-3391}

\begin{abstract}
We derive an analytical expression of a Wigner function that approximately describes the time evolution of the one-dimensional motion of a particle in a nonharmonic potential. Our method involves two exact frame transformations, accounting for both the classical dynamics of the centroid of the initial state and the rotation and squeezing about that trajectory. Subsequently, we employ two crucial approximations, namely the constant-angle and linearized-decoherence approximations, upon which our results rely. These approximations are effective in the regime of wide potentials and small fluctuations, namely potentials that enable spatial expansions orders of magnitude larger than the one of the initial state but that remain smaller compared to the relevant dynamical length scale (e.g., the distance between turning points). Our analytical result elucidates the interplay between classical and quantum physics and the impact of decoherence during nonlinear dynamics. This analytical result is instrumental to designing, optimizing, and understanding proposals using nonlinear dynamics to generate macroscopic quantum states of massive particles.
\end{abstract}		

\maketitle

\section{Introduction}

The one-dimensional position and momentum of a particle is perhaps the most fundamental degree of freedom in physics and of paramount importance in the development of quantum mechanics~\cite{Schrodinger1926,Born1926,Heisenberg1927}. The state of such a continuous-variable degree of freedom can be described by the Wigner function~\cite{Wigner1932, Schleich2001, Case2008}, a phase-space quasi-probability distribution. The Wigner function and its time evolution allow us to study the interplay between classical and quantum mechanics as well as the impact of noise and decoherence~\cite{Zurek1988, Zurek2001, Zurek2007}. Usually, the dynamics of the Wigner function is studied in scenarios of small phase-space areas, i.e., scenarios where the phase-space surface occupied by the initial state is of the same order as the space available to be explored through the dynamics. From a theoretical viewpoint, scenarios with small phase-space areas are advantageous as they circumvent the numerical instabilities associated with the nonlocal and highly oscillatory character of the Wigner equation~\cite{Cabrera2015, RodaLlordes2023_2}, which appear for large coherent expansions due to the presence of arbitrarily high-order derivatives. Experimentally, this is the natural regime for quantum experiments performed with photons~\cite{Deleglise2008, Hofheinz2009, Kirchmair2013} and atomic systems~\cite{Leibfied1996, Fluhmann2020}, whose state, even at near-zero temperatures, is delocalized over scales comparable to the length scale of the confining potential. Motivated by recent experimental progress in controlling the quantum center-of-mass motion of a single nanoparticle~\cite{Gonzalez-Ballestero2021, Delic2020, Magrini2021, Tebbenjohanns2021, Kamba2022, Ranfagni2022, Piotrowski2023, Kamba2023}, which has approximately nine orders of magnitude more mass than a single atom, here we focus on the rather unexplored regime of large-scale quantum dynamics. That is, the dynamics generated when a highly localized phase-space probability distribution evolves in a wide, nonharmonic potential that enables the particle to explore an orders-of-magnitude larger phase-space area.

In this paper, we present a method to obtain an approximated analytical expression for the time evolution of the Wigner function in a nonharmonic potential, in the presence of decoherence, caused by weak coupling to a high-temperature bath~\cite{Caldeira1983, Unruh1989, Breuer2002}, and white-noise fluctuations in both the amplitude and position of the potential~\cite{Gehm1998, Schneider1999, Henkel1999}. Our method implements two exact frame transformations. The first one accounts for the classical trajectory of the centroid of the initial state. The second incorporates the rotation and squeezing around the centroid classical trajectory. After those exact transformations, we perform two approximations, namely the constant-angle and linearized-decoherence approximations. 

On the one hand, the constant-angle approximation is applicable when the local phase-space rotation angle (defined below) around the centroid classical trajectory changes slowly over time. This approximation is expected to hold in the regime of wide potentials, where the relevant length scale of the potential is much larger than the initial spatial extent of the particle's position, thereby allowing large coherent expansions. Importantly, the approximation breaks down close to the turning points, where the quantum state compresses and the local phase-space rotation angle changes rapidly. On the other hand, the linearized-decoherence approximation is accurate in the limit of small fluctuations, where the de Broglie wavelength of the quantum state remains significantly smaller than the relevant length scale of the potential. Both, the breakdown of the approximation near the turning points and the small de Broglie wavelength evoke properties reminiscent of the so-called semiclassical methods~\cite{Berry1972, Berry1977, Keller1985, Zworski2022}, such as the Wentzel–Kramers–Brillouin approximation, that study the asymptotic limit $\hbar\to0$ of the Schr\"odinger equation. Among semiclassical methods, we highlight the connection of our method with the Gaussian wave packet dynamics developed by Heller and co-workers in a series of works~\cite{Heller1975, Heller1976, Heller1977, Davis1984, Littlejohn1986, Huber1988}. This method relies on the so-called \textit{thawed Gaussian approximation}~\cite{Heller2018}, in which the potential is replaced by its quadratic expansion along the classical trajectory of the wave packet. Consequently, our method reproduces this approximation in the limit in which all nonlinear and decoherence effects are disregarded. However, the precise connection between the method discussed here and previously studied semiclassical methods remains unclear, and we leave it for future investigation.

Here, we show that the analytical expression obtained with this method provides an excellent approximation to the generated nonlinear open dynamics in the regime of wide potentials and small fluctuations. Our results are timely as they synergize with recently developed numerical tools~\cite{RodaLlordes2023_2} that allow to design, optimize, and understand protocols where a quantum ground-state-cooled nanoparticle in a wide nonharmonic potential rapidly evolves into a macroscopic quantum superposition state~\cite{RodaLlordes2023}.

The rest of the paper is organized as follows. In Sec.~\ref{sec:quantum_dynamics_wigner}, we derive an analytical approach using Wigner analysis to describe the dynamics of a particle in a nonharmonic potential. We define the regime in which a key approximation, referred to as the constant-angle approximation, is used to integrate the nonlinear dynamics, thereby providing an analytical form of the time-evolved Wigner function. In Sec.~\ref{sec:example}, we apply our analytical method to the example of a particle evolving in a wide double-well potential. We explicitly show how our analytical approach reproduces the numerical results obtained using the split-method operator~\cite{Leforestier1991} and the numerical method presented in~\cite{RodaLlordes2023_2}. We draw our conclusions in Sec.~\ref{sec:conclusions} and provide further details of our analysis in the Appendix.

\section{Wigner Analysis of Particle Nonlinear Dynamics}\label{sec:quantum_dynamics_wigner}

In this Section, we provide an analytical treatment of the one-dimensional dynamics of a particle evolving in a nonharmonic potential in the presence of decoherence. Since the classical equations of motion associated with a nonharmonic potential are nonlinear, we say that the particle undergoes nonlinear dynamics. Initially (Sec.~\ref{subsec:description}), we will delve into the detailed description of the dynamical problem under study, presenting both the Liouville-von Neumann equation and the equation of motion of the Wigner function. Subsequently, we will conduct two frame transformations that are exact: the classical centroid frame transformation (Sec.~\ref{subsec:centroid}), and the Gaussian frame transformation (Sec.~\ref{subsec:gaussian}). These two transformations yield a valuable and exact reformulation of the dynamical problem. Following this, we will perform two approximations (Sec.~\ref{subsec:approx}): the constant-angle approximation and the linearized-decoherence approximation. These will lead to an easily evaluable expression of the time-evolved Wigner function. Finally (Sec.~\ref{subsec:validity}), we will discuss the regime where we expect the constant-angle approximation to provide an accurate description of the nonlinear particle dynamics. 

Throughout the section, we will use the following nomenclature for the different types of dynamics. First, the term \textit{linear dynamics} refers to the dynamics generated by a Hamiltonian whose associated classical equations of motion are linear. That is, whose solutions are closed under linear combinations. This implies that the corresponding Hamiltonian is at most quadratic in the position and momentum variables. The term \textit{coherent dynamics} refers to the dynamics that preserves the coherence and purity of the state over time. In the wave function picture, those are the dynamics generated by the Schr\"odinger equation. Finally, we use the term \textit{Gaussian dynamics} to refer to the dynamics that preserves the Gaussianity of the state. Namely, the combination of linear dynamics plus diffusion-like dissipation. Finally, given the definitions above, we use interchangeably the terms linear dynamics and coherent Gaussian dynamics.

\subsection{Description of the dynamical problem} \label{subsec:description}

We consider the one-dimensional dynamics of a particle with mass $\mass$, where the position and momentum operators are given by $\posunitsop$ and $\momunitsop$, respectively, and they fulfill the canonical commutation rule $[\posunitsop,\momunitsop]=i \hbar $. The particle evolves under a potential denoted by $\potential(\posunitsop)$, and in the presence of decoherence. The time evolution of the particle's state, specified by its density operator $\state(t)$, is governed by the Liouville-von Neumann equation:
\begin{align}
\frac{\partial \state}{\partial t}(t) = \frac{1}{i\hbar}\left[\frac{\momunitsop^2 }{2\mass} +\potential(\posunitsop),\state(t)\right] + \dissipator[\state(t)],\label{eq:liouville_vonNeumann_equation}
\end{align}
Here, $\dissipator[\cdot]$ represents the decoherence superoperator, which we specify below. We are interested in studying the evolution of a particle that is initially cooled to a low-temperature thermal state in a harmonic potential of frequency $\freqtrap$ (for example, an optical trap). At zero temperature, the particle has zero-point fluctuations, with $\zppos = \sqrt{\hbar/(2\mass\freqtrap)}$ in position, and $\zpmom = \hbar/(2 \zppos)$ in momentum.

Motivated by experiments with levitated nanoparticles in ultra-high vacuum~\cite{Gonzalez-Ballestero2021, Delic2020, Magrini2021, Tebbenjohanns2021, Kamba2022, Ranfagni2022, Piotrowski2023, Kamba2023, Dania2023}, we will model the decoherence as follows. First, we consider position localization decoherence~\cite{Joos1985, Schlosshauer2007, RomeroIsart2011}, which is defined by the dissipator:
\begin{align}
\dissipatorloc[\cdot] = -\frac{\dissrateloc}{2 \zppos^2}[\posunitsop,[\posunitsop,\cdot]],\label{eq:position_localization_noise}
\end{align}
where $\dissrateloc$ is the position localization decoherence rate. This type of decoherence emerges in the weak-coupling high-temperature limit of Brownian motion~\cite{Caldeira1983, Unruh1989, Hu1992, Breuer2002} as well as after performing an ensemble average of a white-noise stochastic force~\cite{vanKampen1976}. This decoherence model mimics the recoil heating due to laser light~\cite{Jain2016, Maurer2022} and the emission of thermal photons~\cite{Schlosshauer2007, Romero-Isart2011, Bateman2014, Agrenius2023}, among other sources of decoherence. The dissipator in $\dissipatorloc[\cdot]$ corresponds to the long-wavelength limit of the decoherence superoperator~\cite{Gallis1990, Romero-Isart2011}, and it overestimates the actual decoherence of the system. Hence, even in the case where the form in Eq.~\eqref{eq:position_localization_noise} is not accurate, it can be used to upper bound the effect of environmental decoherence. Additionally to $\dissipatorloc[\cdot]$, we take into account the decoherence due to fluctuations in the potential's center and its amplitude~\cite{Gehm1998, Schneider1999, Henkel1999}. In particular, we consider the fluctuating potential:
\begin{align}
\fluctuatingpotential(\posunits,t) = (1 + \stochproc_2(t)) \potential(\posunits-\zppos \stochproc_1(t)) \approx
\potential(\posunits) - \stochproc_1(t) \zppos \frac{\partial \potential}{\partial \posunits}(\posunits) + \stochproc_2(t) \potential(\posunits),\label{eq:fluctuating_potential}
\end{align} 
where $\stochproc_1(t)$ and $\stochproc_2(t)$ are zero-average stochastic processes representing fluctuations of the potential's center and amplitude, respectively. In Eq.~(\ref{eq:fluctuating_potential}), we have assumed small potential fluctuations, which justify the use of a Taylor series expansion in each $\stochproc_i(t)$, truncated to the first order. For simplicity, we assume $\stochproc_i(t)$ to be independent Gaussian white noise processes with the correlation function $\cumaverage{\stochproc_i(t)\stochproc_j(t')} = 2\pi\psd_i \delta_{ij} \delta(t-t')$, where $\cumaverage{\cdot}$ denotes the average over trajectories and $\psd_i$ is a measure of the noise amplitude.
The average over these stochastic processes can be computed using the cumulant expansion method to the second order~\cite{vanKampen1974}, and leads to the decoherence superoperator:
\begin{align}
\dissipatorfluc[\cdot] = - \frac{\pi \psd_1 \zppos^2}{ \hbar^2 } [\frac{\partial \potential}{\partial \posunits}(\posunitsop),[\frac{\partial \potential}{\partial \posunits}(\posunitsop), \cdot]] 
- \frac{ \pi \psd_2 }{ \hbar^2 } [\potential(\posunitsop),[\potential(\posunitsop), \cdot]].\label{eq:dissipatior_fluctuations}
\end{align}
Putting everything together, the decoherence model used in this paper is given by $\dissipator[\cdot] = \dissipatorloc[\cdot]+ \dissipatorfluc[\cdot]$. This dissipator can be compactly written as:
\begin{equation} \label{eq:decoh_dissip_expansion}
\dissipator[\cdot] = -\sum_{n,m=1}^\infty \frac{\dissrate_{nm}}{2 \zppos^{n+m}}[\posunitsop^n,[\posunitsop^m,\cdot]] \equiv \sum_{n,m=1}^\infty \dissipator_{nm}[\cdot].
\end{equation}
This form is obtained by performing a Taylor expansion of the potential, i.e., $\potential(\posunits) = \sum_{n=1}^{\infty} (\partial^n \potential/\partial \posunits^n)(0) \posunits^n/n!$. The expression of the decoherence rates $\dissrate_{nm}$ can be obtained by collecting the corresponding terms from $\dissipatorloc[\cdot]$ and $\dissipatorfluc[\cdot]$, and they are explicitly given in App.~\ref{app:dissipator}. It should be noted that the dissipator $\dissipator_{nm}[\cdot]$ for $n$ or $m$ strictly larger than one, which originates from the potential fluctuations $\dissipatorfluc[\cdot]$, generates non-Gaussian dissipative dynamics.

We chose to use the Wigner function formalism \cite{Wigner1932, Schleich2001} to analytically handle the dynamics generated by Eq.~\eqref{eq:liouville_vonNeumann_equation}. Moreover, we use the dimensionless position $\pos = \posunits/\zppos$, momentum $\mom = \momunits/\zpmom$, and time $\timefreq= \freq t$ variables, where $\freq$ is an arbitrary frequency scale associated with the potential $\potential(\posunits)$. The corresponding dimensionless potential in these units is defined as $\varpotential(\pos) \equiv \potential(\pos \zppos)/(\mass \freq^2 \zppos^2)$. In accordance with Eq.~\eqref{eq:liouville_vonNeumann_equation}, the equation of motion for the dimensionless Wigner function $\wigner(\psvec,\timefreq)$ is given by
\begin{equation} \label{eq:wigner_equation}
\frac{\partial \wigner}{\partial \timefreq} (\psvec,\timefreq) = \left( \linc + \linq+ \lind \right) \wigner(\psvec,\timefreq),
\end{equation}
where $\psvec \equiv (\pos,\mom)^\transpose$ denotes a point in the dimensionless phase space.
The first term generates classical (i.e., Liouville) dynamics and is expressed as:
\begin{equation} \label{eq:L_c}
\linc = -\frac{\freqtrap}{\freq} \mom \partialx + \frac{\freq}{\freqtrap} \frac{\partial \varpotential}{\partial \pos} (\pos) \partialp.
\end{equation}
The second term generates genuine quantum dynamics and is written as:
\begin{equation} \label{eq:L_q}
\linq = \frac{\freq}{\freqtrap} \sum_{n=1}^\infty \frac{(-1)^n}{(2n+1)!} \frac{\partial^{2n+1} \varpotential}{\partial \pos^{2n+1}} (\pos)\left(\partialp\right)^{2n+1}.
\end{equation}
Note that $\linc + \linq$ generates the same evolution as the Schrödinger equation and that $\linq = 0$ for up to quadratic potentials.
Lastly, the third term generates decoherence, which in accordance with Eq.~(\ref{eq:decoh_dissip_expansion}), can be written as $\lind = \sum_{n,m=1}^\infty \lindnm $, where:
\begin{equation}
\lindnm = \sum_{k=2}^{n+m} \coeffnmk \, x^{n+m-k} \left(\frac{\partial}{\partial\mom}\right)^{k}.\label{eq:dissipator_wigner_space}
\end{equation}
The specific expression of the coefficients $\coeffnmk$ is provided in App.~\ref{app:dissipator}. Note that while we have focused on the one-dimensional motion of massive particles, our results, especially when written in this dimensionless form, can be applied to describe the nonlinear dynamics of other continuous-variable degrees of freedom (e.g., a single electromagnetic field mode). 

The aim of this paper is to solve Eq.~(\ref{eq:wigner_equation}) for an initial Gaussian state, whose Wigner function can be written as:
\begin{align}
\wigner(\psvec,0) = \Gaussian[\covariance(0)](\psvec-\firstmoms(0)).\label{eq:initial_gaussian}
\end{align}
Here, $\Gaussian[\covariance](\psvec)$ denotes the two-dimensional Gaussian distribution:
\begin{align}
\Gaussian[\covariance](\psvec) \equiv \frac{1}{2\pi \det \covariance} \exp\left(-\frac{\psvec^\transpose \covariance^{-1} \psvec}{2}\right),\label{eq:gaussian_function}
\end{align}
and $\firstmoms(0)=(\average{\hat x}(0),\average{\hat p}(0))^\transpose$ and $\covariance(0)$ refer to the initial mean values and the covariance matrix of $\state(0)$, respectively. The elements $C_{ij}(0)$, where $i,j \in \{\pos,\mom\}$, of the always symmetric covariance matrix are given by:
\begin{equation}
C_{ij}(0) = \frac{1}{2} \average{\hat r_i\hat r_j+\hat r_j\hat r_i}(0)- \average{\hat r_i}(0) \average{\hat r_j}(0).
\end{equation}
For instance, a thermal state with a mean phonon occupation number $\nth$ serves as a particularly relevant initial state. This state corresponds to $\covariance(0) = (2\nth + 1)\identity_2$, where $\identity_2$ represents the two-by-two identity matrix.

In the following section, we will derive an expression for $\wigner(\psvec,\timefreq)$, which approximately solves Eq.~(\ref{eq:wigner_equation}). We obtain this expression by performing two consecutive frame transformations (i.e., the classical centroid frame and the Gaussian frame) and applying a key approximation, namely the constant-angle approximation. We will further simplify this expression using the linearized-decoherence approximation.

\subsection{Classical centroid frame}\label{subsec:centroid}

The first frame transformation aims to remove the dynamics of the mean position and momentum values, effectively moving to a comoving frame that aligns with the approximate centroid of the Wigner function~\cite{Heller1975}. In our Wigner function context, such frame transformations follow from the sometimes called \textit{disentangling theorem}~\cite{Feynman1951, vanKampen1974a}, and are a consequence of the properties of exponential dynamical maps. Essentially, they are equivalent to the transformation between different pictures in standard quantum mechanics. In order to implement the frame transformation, we define $\wignercl(\timefreq) \equiv \mapcl(\timefreq)^{-1} \wigner(\psvec,\timefreq)$, where $\mapcl(\timefreq)$ is a time-dependent phase-space displacement map, expressed as:
\begin{align}
\mapcl(\timefreq) \equiv \exp\left[-\psveccl(\timefreq)^\transpose \nabla \right] = \exp\left[-\poscl(\timefreq)\partialx - \momcl(\timefreq)\partialp\right].\label{eq:classical_trajectory_map}
\end{align}
where $\nabla = (\partial/\partial\pos,\partial/\partial\mom)^\transpose$ is the phase space gradient. In Eq.~\eqref{eq:classical_trajectory_map}, $\psveccl(\timefreq) \equiv (\poscl(\timefreq),\momcl(\timefreq))^\transpose$ represents the classical centroid trajectory, which is the solution to dimensionless Hamilton's equations
\begin{align}
&\frac{\partial \poscl}{\partial \timefreq} (\timefreq) = \frac{\freqtrap}{\freq} \momcl(\timefreq),\label{eq:xcdot}\\
&\frac{\partial \momcl}{\partial \timefreq}(\timefreq) = -\frac{\freq}{\freqtrap} \frac{\partial \varpotential}{\partial x}(\poscl(\timefreq)),\label{eq:pcdot}
\end{align}
with initial conditions given by $\psveccl(0) = (\poscl(0),\momcl(0))^\transpose =(\average{\hat x}(0),\average{\hat p}(0))^\transpose$. The action of the map $\mapcl(\timefreq)$ on an arbitrary function $\testfunc(\psvec)$ translates its phase-space variables as $\mapcl(\timefreq) \testfunc(\psvec) = \testfunc(\psvec-\psveccl(\timefreq))$. In this frame, we describe the dynamics of the Wigner function centered at the classical trajectory that is followed by its initial position and momentum expected value. This frame transformation does not imply any approximation. It will be particularly useful for nonlinear dynamics that maintain a small distance between the vector $\psveccl(\timefreq) = (\poscl(\timefreq),\momcl(\timefreq))^\transpose$ and $\firstmoms(\timefreq)=(\average{\hat x}(\timefreq),\average{\hat p}(\timefreq))^\transpose$ in comparison to the available phase-space dimensions. As we will demonstrate later, this will be the case for wide potentials and small fluctuations.

In this classical centroid frame, one can show that the state $\wignercl(\psvec,\timefreq)$ evolves according to an effective time-dependent potential, represented by
\begin{equation} \label{eq:U_eff}
\varpotentialeff(x,\timefreq) \equiv \sum_{n=2}^\infty \frac{1}{n!}\frac{\partial^n \varpotential}{\partial \pos^n}(\poscl(\timefreq)) x^n.
\end{equation}
This effective potential, at least quadratic in position, exhibits time dependence determined by the local derivatives of the potential along the classical trajectory. Thus, in the comoving frame with the approximate centroid of the Wigner function, the particle experiences a time-dependent potential while traversing the static potential.
More specifically, the evolution equation for $\wignercl(\psvec,\timefreq)$ can be expressed as:
\begin{align}
\frac{\partial \wignercl}{\partial \timefreq} (\psvec,\timefreq) = \left[ \linclc(\timefreq) + \linclq(\timefreq)+ \lincld(\timefreq)\right]\wignercl(\psvec,\timefreq),
\end{align}
Here, $\linclc(\timefreq)$ and $ \linclq(\timefreq)$ are obtained by replacing $\varpotential(x)$ with $\varpotentialeff(\pos,\timefreq)$ in Eq.~(\ref{eq:L_c}) and Eq.~(\ref{eq:L_q}), respectively. The generator of decoherence also becomes time-dependent and is given by $\lincld(\timefreq) \equiv \mapcl(\timefreq)^{-1} \lind \mapcl(\timefreq) = \sum_{n,m=1}^\infty \lincldnm(\timefreq)$, with
\begin{equation}
\lincldnm(\timefreq) = \sum_{k=2}^{n+m} \coeffnmk \, \left[x+\poscl(\timefreq)\right]^{n+m-k} \left(\frac{\partial}{\partial\mom}\right)^{k}.
\end{equation}

\subsection{Gaussian frame}\label{subsec:gaussian}

In the centroid frame, the coherent dynamics are generated by the sum of $\linclc(\timefreq)+\linclq(\timefreq)$. The second frame transformation aims to remove the coherent Gaussian dynamics generated by the harmonic part of the effective potential Eq.~(\ref{eq:U_eff}), which we define as:
\begin{equation} \label{eq:U_g}
\varpotentialgauss(\pos,\timefreq) \equiv \frac{1}{2} \funct(\timefreq) \pos^2 =  \frac{1}{2}\frac{\partial^2 \varpotential}{\partial x^2}(\poscl(\timefreq)) \pos^2 .
\end{equation}
Here, $\funct(\timefreq)$ represents an effective (dimensionless) spring constant, which can be either positive or negative. To remove the coherent Gaussian dynamics, we define $\wignerhar(\psvec,\timefreq) \equiv \maphar(\timefreq)^{-1} \wignercl(\psvec,\timefreq)$, where $\maphar(\timefreq)$ is given by:
\begin{align}
\maphar(\timefreq) \equiv \toexp\left[\int_0^\timefreq d\timefreq' \linclG(\timefreq')\right],\label{eq:linear_evolution_frame}
\end{align}
where $\toexp(\cdot)$ represents the time-ordered exponential and $\linclG(\timefreq)$ is the generator of coherent Gaussian dynamics in the centroid frame, defined as:
\begin{align}
&\linclG(\timefreq) = -\frac{\freqtrap}{\freq} p \partialx + \frac{\freq}{\freqtrap} \frac{\partial \varpotentialgauss}{\partial x}(\pos,\timefreq) \partialp.
\end{align}
The generator of the coherent non-Gaussian dynamics, attributed to the nonharmonic terms of the potential, is given by the complementary generator $\linclnG(\timefreq) \equiv \linclc (\timefreq) + \linclq(\timefreq) - \linclG(\timefreq)$. While assessing if a given state of motion qualifies as a quantum non-Gaussian state lies beyond the scope of this paper, we remark that it is possible to find sufficient criteria based on the observation of squeezing in nonlinear variables~\cite{Moore2022}.

The map $\maphar(\timefreq)$ generates coherent Gaussian physics, namely squeezing and phase-space rotations. Its action on an arbitrary function $\testfunc(\psvec)$ is $\maphar(\timefreq)\testfunc(\psvec) = \testfunc(\propagator(\timefreq)^{-1}\psvec)$, where $\propagator(\timefreq)$ is a symplectic matrix. The matrix $\propagator(\timefreq)$ is obtained as the solution to the differential equation
\begin{align}
\frac{\partial\propagator}{\partial \timefreq}(\timefreq) = \begin{pmatrix} 0 & \freqtrap/\freq \\ -(\freq/\freqtrap) \funct(\timefreq) & 0 \end{pmatrix} \propagator(\timefreq),\label{eq:evolution_S}
\end{align}
with the initial condition $\propagator(0) = \identity_2$. One can verify that $\text{det}[\propagator(\timefreq)] = 1$ for all $\timefreq$. Therefore, denoting $\propagatorcomp_{ij}(\timefreq)$ with $i,j \in \{\pos,\mom\}$ the components of $\propagator(\timefreq)$, its inverse is given by
\begin{align}
\propagator(\timefreq)^{-1} = \begin{pmatrix}
\propagatorcomp_{pp}(\timefreq) & -\propagatorcomp_{xp}(\timefreq)\\
-\propagatorcomp_{px}(\timefreq) & \propagatorcomp_{xx}(\timefreq)
\end{pmatrix}.
\end{align}
Importantly, the action of the map $\maphar(\timefreq)$ on the phase space variable $\pos$ is given by
\begin{align}
\maphar(\timefreq)^{-1} \pos \maphar(\timefreq) \equiv \eta(\timefreq) \pos_{\quadratureangle(\timefreq)} = \inflation(\timefreq)[\cos(\quadratureangle(\timefreq)) \pos + \sin(\quadratureangle(\timefreq))\mom],\label{eq:local_rotation_angle}
\end{align}
where $\psvec_{\quadratureangle(\timefreq)} \equiv \rotation[\quadratureangle(\timefreq)] \psvec =(\pos_{\quadratureangle(\timefreq)} ,\mom_{\quadratureangle(\timefreq)} )^\transpose $ are rotated phase-space variables according to the rotation matrix
\begin{align}
\rotation[\quadratureangle(\timefreq)] \equiv \begin{pmatrix} \cos[\quadratureangle(\timefreq)] & \sin[\quadratureangle(\timefreq)]\\
-\sin[\quadratureangle(\timefreq)] & \cos[\quadratureangle(\timefreq)]
\end{pmatrix},\label{eq:rotation}
\end{align}
and we have defined the two key variables
\begin{align}
&\inflation(\timefreq) \equiv \sqrt{\propagatorcomp_{xx}(\timefreq)^2 + \propagatorcomp_{xp}(\timefreq)^2},\label{eq:coherent_inflation}\\
&\tan [\quadratureangle(\timefreq) ]\equiv \frac{\propagatorcomp_{xp}(\timefreq)}{\propagatorcomp_{xx}(\timefreq)}\label{eq:quadrature_angle}.
\end{align}
From Eq.~\eqref{eq:local_rotation_angle}, the angle $\quadratureangle(\timefreq)$ can be regarded as the local rotation angle of the position quadrature in phase space. Similar transformation rules can be derived for $\partial/\partial \mom$, involving the same functions $\inflation(\timefreq)$ and $\quadratureangle(\timefreq)$, and for $\mom$ and $\partial/\partial\pos$ using similar functions that combine $\propagatorcomp_{\mom\mom}(\timefreq)$ and $\propagatorcomp_{\mom\pos}(\timefreq)$ instead of $\propagatorcomp_{\pos\pos}(\timefreq)$ and $\propagatorcomp_{\pos\mom}(\timefreq)$. In summary, the map $\maphar(\timefreq)$ generates time-dependent phase-space rotations with an angle $\quadratureangle(\timefreq)$ and time-dependent squeezing with a squeezing parameter $\inflation(\timefreq)$.

In the centroid and Gaussian frame, the evolution equation of $\wignerhar(\psvec,\timefreq)$ is given by
\begin{align} \label{eq:wigner_equation_gaussian}
\frac{\partial \wignerhar}{\partial \timefreq} (\psvec,\timefreq) = \left[ \linharnG(\timefreq) + \linhard(\timefreq)\right]\wignerhar(\psvec,\timefreq),
\end{align}
where $\linharnG(\timefreq) \equiv \maphar(\timefreq)^{-1} \linclnG (\timefreq)\maphar(\timefreq)$ and $\linhard(\timefreq) \equiv \maphar(\timefreq)^{-1} \lincld (\timefreq) \maphar(\timefreq) $ that also can be expanded as $\linhard(\timefreq) =  \sum_{n,m=1}^\infty \linhardnm$. The explicit form of the generator of coherent non-Gaussian dynamics is given by
\begin{align}
\linharnG(\timefreq) =& \sum_{n=2}^\infty \betas_{n+1}(\timefreq) \pos_{\quadratureangle(\timefreq)}^n \frac{\partial}{\partial \mom_{\quadratureangle(\timefreq)}} \nonumber\\
&\hspace{2cm}+ \sum_{\substack{n=1 \\ m=0}}^\infty (-1)^n \binom{2n+m}{m} \frac{\betas_{2n+m+1}(\timefreq)}{2n+1} \pos_{\quadratureangle(\timefreq)}^m \left(\frac{\partial}{\partial \mom_{\quadratureangle(\timefreq)}}\right)^{2n+1},
\end{align}
where the first and second terms account for classical and quantum non-Gaussian dynamics, respectively, and we have defined
\begin{align}
&\betas_n(\timefreq) \equiv \frac{\freq}{\freqtrap} \frac{1}{(n-1)!} \frac{\partial^n \varpotential}{\partial \pos^n} (\poscl(\timefreq)) \inflation(\timefreq)^{n}.\label{eq:betas}
\end{align}
Note that the magnitude of the parameters $\beta_n(\timefreq)$ will determine whether non-Gaussian dynamics is generated. This shows that large squeezing, that is, a large value of $\inflation(\timefreq)$, enhances the effect of the nonharmonicities in the potential~\cite{Rosiek2023}. The specific form of the generators of decoherence is given by
\begin{align} 
\linhardnm(\timefreq) = \sum_{k=2}^{n+m} \coeffnmk \left[\inflation(\timefreq)\pos_{\quadratureangle(\timefreq)}+\poscl(\timefreq)\right]^{n+m-k} \inflation(\timefreq)^k \left(\frac{\partial}{\partial \mom_{\quadratureangle(\timefreq)}}\right)^{k}.\label{eq:L_decoh_gaussian}
\end{align}
Note that decoherence is also enhanced by squeezing.

We emphasize that up to this point, the analytical approach is exact; that is, no approximations have been made. We have, however, singled out the challenging part of solving a nonlinear open quantum dynamical problem, which is to integrate Eq.~\eqref{eq:wigner_equation_gaussian} with the initial condition given by $\wignerhar(\psvec,0) = \maphar(0)^{-1}\mapcl(0)^{-1} \wigner(\psvec,0)$. If one obtains $\wignerhar(\psvec,\timefreq)$, the exact solution of Eq.~\eqref{eq:wigner_equation} is given by
\begin{align}
\wigner(\psvec,\timefreq) = \mapcl(\timefreq)\maphar(\timefreq)\wignerhar(\psvec,\timefreq).\label{eq:exact_solution}
\end{align}
Interestingly, replacing $\wignerhar(\psvec,\timefreq)$ by $\wignerhar(\psvec,0)$ in Eq.~\eqref{eq:exact_solution}, one obtains approximate Gaussian dynamics that are equivalent to performing the so-called Gaussian thawed approximation~\cite{Heller1975, Heller2018} but in the Wigner, instead of the wave function, representation. Note that such a replacement is equivalent to ignoring all the nonlinear and decoherence effects. Moreover, as we show below, the exact reformulation of the problem presented here allows us to identify a crucial approximation, yielding an approximated Wigner function $\wignerharapprox(\psvec,\timefreq)$, that incorporates nonlinear effects while being significantly easier to calculate than its exact counterpart.

\subsection{Constant-angle and linearized-decoherence approximation}\label{subsec:approx}

Ultimately, the reason a closed-form solution for $\wignerhar(\psvec,\timefreq)$ is not possible stems from the fact that the angle $\quadratureangle(\timefreq)$ changes over time. This results in the generators $\linharnG(\timefreq)$ and $\linhard(\timefreq)$ being noncommutative with themselves at different times, and the solution of the Wigner equation being a time-ordered instead of a simple exponential map. As we shall see, in the regime of wide potentials and small fluctuations, where $\poscl(\timefreq) \gg \inflation(\timefreq) \gg 1$, it becomes feasible to approximate the angle $\quadratureangle(\timefreq)$ as a piecewise constant function. This approximation significantly simplifies the integration of the evolution equation for the Wigner function. Let us focus on the simplest case, where $\quadratureangle(\timefreq)$ can be replaced by a single constant angle $\constantangle$ over the integration regime of interest, while the more general case is discussed in App.~\ref{app:constant_angle_approximation}. In this scenario, the generator of Eq.~\eqref{eq:wigner_equation_gaussian} commutes with itself at different times, and furthermore, all its summands also commute with each other. We will first implement this approximation and then, in the subsequent subsection, discuss the regime where we expect it to provide an accurate description of the particle's nonlinear dynamics.

Within the constant-angle approximation, the approximated solution of $\wignerhar(\psvec,\timefreq)$ is given by an exponential map (rather than a time-ordered exponential one, see Ch.~2 of~\cite{Mukamel1999}), which is time-dependent and can be factorized. This can be expressed as follows:
\begin{equation}
\wignerhar(\psvec,\timefreq) \approx  \wignerharapprox(\psvec,\timefreq) = \mapng (\timefreq) \mapnoisehar(\timefreq) \wignerhar(\psvec,0).
\end{equation}
Here, $\mapng (\timefreq)$ generates coherent non-Gaussian dynamics and is given by the exponential map $\mapng (\timefreq) = \exp [\intlin_\nongaussian(\timefreq)]$, where
\begin{equation}
\intlin_\nongaussian(\timefreq) \equiv \sum_{n=2}^\infty \kappas_{n+1}(\timefreq) \pos_{\constantangle}^n \frac{\partial}{\partial \mom_{\constantangle}}
+\sum_{\substack{n=1\\m=0}}^\infty (-1)^n \binom{2n+m}{m} \frac{\kappas_{2n+m+1}(\timefreq)}{2n+1} \pos_{\constantangle}^m \left(\frac{\partial}{\partial \mom_{\constantangle}}\right)^{2n+1}.\label{eq:map_approx_ng} 
\end{equation}
The first and second terms account for classical and quantum non-Gaussian dynamics, respectively. The time-dependent coefficients quantifying the strength of the generator of these coherent non-Gaussian dynamics are given by
\begin{align}
\kappas_n(\timefreq) \equiv \int_0^\timefreq d\timefreq' \betas_n(\timefreq') = \frac{\freq}{\freqtrap}\frac{1}{(n-1)!} \int_0^\timefreq d\timefreq' \frac{\partial^n \varpotential }{\partial x^n}(\poscl(\timefreq'))\inflation(\timefreq')^{n}.\label{eq:kappas}
\end{align}
Similarly, \old{the decoherence in the Gaussian frame is generated by}  \new{$\mapnoisehar(\timefreq) \equiv \exp[\intlind(\timefreq)]$ accounts for the decoherence in the Gaussian frame, where}
\begin{align} 
\intlind(\timefreq) \equiv \int_0^\timefreq d\timefreq' \sum_{n,m=1}^\infty \sum_{k=2}^{n+m} \coeffnmk \, \poscl(\timefreq')^{n+m} \left[1 + \frac{\inflation(\timefreq')}{\poscl(\timefreq')}\pos_{\constantangle}\right]^{n+m-k} \left[\frac{\inflation(\timefreq')} {\poscl(\timefreq')} \right]^k
\left(\frac{\partial}{\partial \mom_{\constantangle}}\right)^{k}.\label{eq:noise_map_general}
\end{align}

Note that by using the constant-angle approximation, the solution to the dynamics in Wigner space has been significantly simplified. This is because one has replaced the integration of a complicated partial differential equation with the more efficient integration of a set of time-dependent functions. 

In addition to the constant-angle approximation, a second approximation can be performed on the decoherence map under the assumption of small fluctuations. This means that the following condition is met:
\begin{equation} \label{eq:small_fluctuations}
\frac{\inflation(\timefreq)}{\poscl(\timefreq) } \ll 1.
\end{equation}
This is the regime in which the position fluctuations of the particle are smaller than its mean value. Even in the case where $\poscl(\timefreq) = 0$, the linearization of the decoherence generator may still be possible, albeit for different underlying reasons (see App.~\ref{app:underlying_reason}). Under this regime, one can perform the linearized-decoherence approximation, which consists of expanding Eq.~(\ref{eq:noise_map_general}) in powers of $\inflation(\timefreq)/\poscl(\timefreq)$ and keeping the lowest order terms up to the second order. The lowest-order terms can be written compactly as follows:
\begin{equation}
\intlind\new{(\tau)} \approx \intlindapprox(\timefreq)\equiv \frac{\sigmablurring(\timefreq)^2}{2} \left(\frac{\partial}{\partial\mom_{\constantangle}}\right)^2.
\end{equation}
The time-dependent parameter $\sigmablurring(\timefreq)$ quantifies the strength of the accumulated decoherence and is given by 
\begin{align}
\sigmablurring(\timefreq)^2 \equiv 4 \int_0^\timefreq d\timefreq' \frac{\dissrateeff(\timefreq')}{\freq} \inflation(\timefreq')^2.\label{eq:blurring_distance}
\end{align}
The effective time-dependent decoherence rate is given by $\dissrateeff(\timefreq) \equiv \dissrateloc+ \dissratefluc(\timefreq)$, with $\dissrateloc$ defined below Eq.~\eqref{eq:position_localization_noise} and with the contribution arising from the fluctuations of the potential
\begin{equation} \label{eq:Gamma_fluc}
\frac{\dissratefluc(\timefreq)}{\freqtrap} = \frac{\pi\freq^4 \psd_1}{2\freqtrap^3} \left( \left[\frac{\partial^2 \varpotential}{\partial x^2}(\poscl(\timefreq))\right]^2 +\frac{\psd_2}{\psd_1} \left[\frac{\partial \varpotential}{\partial x}(\poscl(\timefreq)) \right]^2 \right).
\end{equation}
Note that this approximation corresponds to replacing the decoherence generator $\linhard(\timefreq) = \sum_{n,m=1}^\infty \linhardnm(\timefreq)$ in Eq.~(\ref{eq:L_decoh_gaussian}), by its linear approximation defined by
\begin{align}
\linhard(\timefreq) \approx \frac{2\dissrateeff(\timefreq)}{\freq} \inflation(\timefreq)^2 \left(\frac{\partial}{\partial\mom_{\constantangle}}\right)^2.\label{eq:linearized_decoherence_gaussian}
\end{align}

Putting everything together, the constant-angle and linearized-decoherence approximations allow us to obtain an approximate solution to the open nonlinear quantum mechanical problem described by Eq.~(\ref{eq:wigner_equation}). This can be represented as follows:
\begin{align} 
\wignerapprox(\psvec,\timefreq) &= \mapcl(\timefreq)\maphar(\timefreq) \mapng (\timefreq) \new{\mapnoiseharapprox}(\timefreq)\wignerhar(\psvec,0) \label{eq:W_a_1}\\
&= \new{\mapnoiseapprox}(\timefreq) \mapcl(\timefreq) \maphar(\timefreq) \mapng(\timefreq) \wignerhar(\psvec,0) \label{eq:W_a},
\end{align}
Here, we have defined $\new{\mapnoiseharapprox}(\timefreq) \equiv \exp [\intlindapprox(\timefreq)]$ and $\new{\mapnoiseapprox}(\timefreq) \equiv \maphar(\timefreq) \new{\mapnoiseharapprox} (\timefreq) \maphar(\timefreq)^{-1}$. Also, recall that $\wignerhar(\psvec,0) = \maphar(0)^{-1}\mapcl(0)^{-1} \wigner(\psvec,0)$. In the second equation, we have used the fact that $\mapnoisehar(\timefreq)$ commutes with $\mapng (\timefreq)$ and $\mapnoise(\timefreq)$ with $\mapcl(\timefreq)$. The expression in the second equation is particularly convenient as the decoherence map is applied to the coherently evolved state. In this case, the decoherence map is given by:
\begin{align}
\new{\mapnoiseapprox}(\timefreq) \equiv \exp\left[\frac{\psgrad^\transpose \Sigmablurring(\timefreq) \psgrad}{2}\right],\label{eq:noise_map}
\end{align}
where
\begin{align}
\Sigmablurring(\timefreq) = \sigmablurring(\timefreq)^2 \propagator(\timefreq) \rotation(\constantangle)^\transpose (\etwo \etwo^\transpose) \rotation(\constantangle) \propagator(\timefreq)^\transpose \label{eq:noise_matrix}
\end{align}
is the blurring covariance matrix and $\etwo = (0,1)^\transpose$ is a unit vector along the momentum component. The action of the decoherence map $\new{\mapnoiseapprox}(\timefreq)$ corresponds to a convolution with a Gaussian distribution of covariance matrix $\Sigmablurring(\timefreq)$, namely:
\begin{align}
\new{\mapnoiseapprox}(\timefreq) \testfunc(\psvec) = \int d\psvec' \Gaussian[\Sigmablurring(\timefreq)](\psvec-\psvec') \testfunc(\psvec').\label{eq:convolution}
\end{align}
Here, the Gaussian distribution $\Gaussian[\covariance](\psvec)$ is defined as per Eq.~\eqref{eq:gaussian_function}. Physically, $\sigmablurring(\timefreq)$ determines the smallest length scale that remains unaffected by decoherence and is sometimes referred to as the blurring distance~\cite{Romero-Isart2017, Pino2018, Neumeier2024}.

We can now use the expression $\wignerapprox(\psvec,\timefreq)$ in Eq.~\eqref{eq:W_a} to calculate the first and second moments. As shown in App.~\ref{subapp:first_moms} and~\ref{subapp:second_moms}, the first moments $\firstmoms_\approximated(\timefreq)$ can be expressed as
\begin{align}
\firstmoms_\approximated(\timefreq) = \psveccl(\timefreq) -\propagator(\timefreq)\rotation(\constantangle)^\transpose \etwo \sum_{n\geq 3} \kappas_n(t) \average{\hat \pos_\constantangle^{n-1}}^\gaussianframe(0).\label{eq:first_moments_nG}
\end{align}
where $\average{\cdot}^{\gaussianframe}(\timefreq)$ is the average in the Gaussian frame; that is, computed with $\wignerhar(\psvec,\timefreq)$. Note that the first moments are not affected by decoherence. The second moments, namely the covariance matrix evaluated using $\wignerapprox(\psvec,\timefreq)$, can be expressed as
\begin{widetext}
\begin{align}
\covariance_\approximated(\timefreq) &= \Sigmablurring(\timefreq)+\propagator(\timefreq)\rotation(\constantangle)^\transpose\left[\covariance(0) -\sum_{n\geq 3} \kappas_n(t) \begin{pmatrix}
0 & \average{\hat \pos_\constantangle^n}^\gaussianframe(0)\\
\average{\hat \pos_\constantangle^n}^\gaussianframe(0) & \average{\hat \pos_\constantangle^{n-1} \hat \mom_\constantangle}^\gaussianframe(0)
\end{pmatrix}\right.\nonumber\\
\hspace{-10pt}+&\left. \sum_{n,m \geq 3} \kappa_n(t) \kappa_m(t) \begin{pmatrix}
0 & 0\\
0 & \average{\hat \pos_\constantangle^{n+m-2}}^\gaussianframe(0) - \average{\hat \pos_\constantangle^{n-1}}^\gaussianframe(0)\average{\hat \pos_\constantangle^{m-1}}^\gaussianframe(0)\end{pmatrix}\right]\rotation(\constantangle)\propagator(\timefreq)^\transpose .\label{eq:second_moments_nG}
\end{align}
\end{widetext}
\noindent Note that decoherence appears in the covariance matrix through the addition of the blurring covariance matrix defined in Eq.~\eqref{eq:noise_matrix}. Nonlinear dynamics is manifested by the dependence of $\covariance_\approximated(\timefreq)$ on the initial moments higher than the second \cite{Weiss2019}. In the next section, we argue that the accuracy of $\firstmoms_\approximated(\timefreq)$ and $\covariance_\approximated(\timefreq)$ in approximating $\firstmoms(\timefreq)$ and $\covariance(\timefreq)$ is a useful measure of the merit of our analytical approach.

\subsection{Validity of the constant-angle approximation}\label{subsec:validity}

Our analytical approach relies on the \textit{constant-angle} and \textit{linearized-decoherence} approximations, which lead to the expression $\wignerapprox(\psvec,\timefreq)$, given in Eq.~\eqref{eq:W_a}. We note that the linearized-decoherence approximation, performed after the constant-angle approximation, is convenient but not essential, and its validity is well characterized by the small-fluctuations condition in Eq.~\eqref{eq:small_fluctuations}. Regarding the constant-angle approximation, we need to compare how closely $\wignerapprox(\psvec,\timefreq)$ approximates the solution $\wigner(\psvec,\timefreq)$. One can show that an upper bound for the norm of the difference of the exact and approximated Wigner functions is proportional to the function
\begin{align} 
\chi(\timefreq;\constantangle) \equiv \sum_{n \geq 3} \int_0^\timefreq d\timefreq'  |\betas_n(\timefreq')| |\quadratureangle(\timefreq') - \constantangle|.\label{eq:chi}
\end{align}
Indeed, this quantity is minimized if the angle $\quadratureangle(\timefreq)$ is constant and close to $\phi$ on the time scales where the nonharmonicities are large, namely when the $\betas_n(\timefreq)$ are large. In this context, from the definition of the angle $\quadratureangle(\timefreq)$ in Eq.~\eqref{eq:quadrature_angle}, it follows that
\begin{align}
\frac{d\quadratureangle}{d\timefreq} (\timefreq) = \frac{\freqtrap/\freq}{\inflation(\timefreq)^2}\label{eq:angle_derivative},
\end{align}
indicating that a large delocalization $\inflation(\timefreq) \gg 1$, reduces the rate of change of the angle. Together with the small-fluctuations condition Eq.~\eqref{eq:small_fluctuations} required for the linearized-decoherence approximation, our analytical approach requires the regime $\poscl(\timefreq) \gg \inflation(\timefreq)\gg1$ in the time scales where the generators of nonlinear dynamics cannot be neglected.

Ultimately, the best way to quantify the performance of our analytical approach is to compare $\wignerapprox(\psvec,\timefreq)$ with the numerically evaluated $\wigner(\psvec,\timefreq)$. In this case, we suggest comparing observables rather than the overlap between the two states. The reason is that large-scale dynamics generates states with very small phase-space features, even sub-Planckian~\cite{Zurek2001, RodaLlordes2023_2} that can immediately provide a very low overlap if not captured accurately. These small features can nevertheless be irrelevant in capturing the observables of interest and may immediately disappear in the presence of unavoidable sources of noise and decoherence. Therefore, we suggest quantifying the performance of the constant-angle approximation by comparing global observable properties of the state such as their first and second moments. This can be done by evaluating their relative error, namely,
\begin{align} \label{eq:epsilon_1}
\epsilon_1(\timefreq) \equiv \frac{2 \norm{\firstmoms(\timefreq)-\firstmoms_\textit{a}(\timefreq)}}{\norm{\firstmoms(\timefreq)}+\norm{\firstmoms_\textit{a}(\timefreq)}},
\end{align}
with the vector norm $\norm{\textbf{y}} = \sqrt{\textbf{y}^\dagger \textbf{y}}$, and
\begin{align} \label{eq:epsilon_2}
\epsilon_2(\timefreq) \equiv \frac{2 \norm{\covariance(\timefreq)-\covariance_\textit{a}(\timefreq)}_2}{\norm{\covariance(\timefreq)}_2+\norm{\covariance_\textit{a}(\timefreq)}_2},
\end{align}
where $\norm{\textbf{A}}_2= \sqrt{\text{tr}(\textbf{A}^\dagger \textbf{A})}$ is the Hilbert-Schmidt norm. In summary, we recommend evaluating $\chi(\timefreq;\constantangle)$, $\epsilon_1(\timefreq)$ and $\epsilon_2(\timefreq)$ to quantify the performance of the constant-angle approximation, something we will do in the example shown in the following section.

Finally, let us emphasize that around a turning point, say $\taumax$, the state is expected to compress, namely to minimize $\inflation(\timefreq)$. Hence, according to Eq.~\eqref{eq:angle_derivative}, around a turning point the angle $\quadratureangle(\timefreq)$ will not change slowly. By performing a Taylor expansion of $\quadratureangle(\timefreq)$ around the time $\taumax$, recall Eq.~\eqref{eq:coherent_inflation}, it follows that the time derivative of the angle around a compression time has a Lorentzian shape
\begin{align}
\frac{d\quadratureangle}{d\timefreq} (\timefreq) \approx \frac{\gamma(\taumax)}{\gamma(\taumax)^2 + (\timefreq-\taumax)^2},
\end{align}
with width $\gamma(\taumax) \equiv (\freq/\freqtrap)\inflation(\taumax)^2$. Therefore, by integrating this Lorentzian, we see that around a turning point, the angle should change approximately by $\pi$. This implies that if the constant-angle approximation is performed with an angle $\constantangle$ during the time scale before a turning point, a second angle around $\constantangle + \pi$ will need to be employed after the turning point. Note that this phase change across a turning point is in agreement with the change in the Maslov index~\cite{Maslov2001} that often appears in semiclassical methods. We will show and use this explicitly in the example provided in the following section.

\section{Example: dynamics in a wide double-well potential}\label{sec:example}

Let's demonstrate how the formalism developed in the previous section can be applied to describe the nonlinear dynamics of a particle using a relevant example. Inspired by a recent proposal to prepare a macroscopic quantum superposition of a levitated nanoparticle via the dynamics in a wide nonharmonic potential~\cite{RodaLlordes2023}, we will focus on the double-well potential given by:
\begin{align}
\potentialdw(\posunits) = \frac{1}{2}\mass \freq^2 \left(-\posunits^2 + \frac{\posunits^4}{2\dwminunits^2}\right).
\end{align}
This potential is characterized by the frequency $\freq$ and the distance scale $\dwminunits$. To adapt the results of the previous section, we will employ the dimensionless potential:
\begin{align}
&\varpotentialdw(\pos) = \frac{1}{2}\left(-\pos^2 + \frac{\pos^4}{2 \dwmin^2}\right),\label{eq:doublewell_potential}
\end{align}
where $\dwminunits = \dwmin \zppos$. We will consider the wide-potential regime defined by $\dwmin \gg 1$. We consider the initial condition given by a thermal state of a harmonic potential of frequency $\freqtrap$ centered at $x=0$ with phonon mean number occupation $\bar n$ and mean position and momentum values denoted by $\firstmoms(0)=(\average{\hat x}(0),\average{\hat p}(0))^\transpose = (\pos_\textrm{s},0)^\transpose$, where $\dwmin \gg \pos_\textrm{s} \gg 1$, such that the dynamics is constrained to the right side of the double-well potential. Whenever necessary, we will utilize the parameters listed in Table~\ref{tab:parameters}, which correspond to the size L dynamics defined in~\cite{RodaLlordes2023}.

\begin{table}[b]
    \centering
    \begin{tabular}{|r||c|c|c|c|} 
    \hline
         Parameter: & $\freq/\freqtrap$ & $\dwmin$ & $\pos_\textrm{s}/\dwmin$ & $\;\nth\;$ \\ \hline
         Value: & $10^{-2}$ & $10^4$ & $10^{-1}$ & 0 \\ \hline
    \end{tabular}
    \caption{Parameters used in the double-well potential example.}
    \label{tab:parameters}
\end{table}

\begin{figure}
    \centering
    \includegraphics[width=\linewidth]{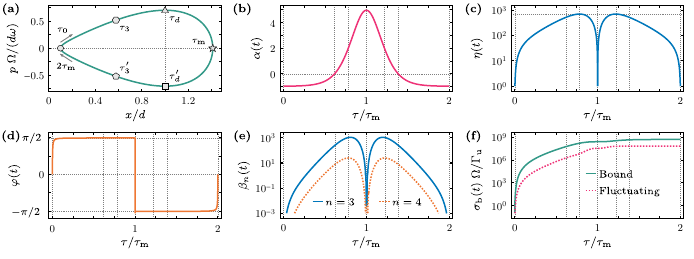}
    \caption{Relevant functions for our analytical approach computed for the parameters in Table~\ref{tab:parameters}.
    Panel (a) shows the classical trajectory $\psveccl(\timefreq)$ in units of $d$. 
    Panels (b),(c),(d),(e) and (f) show, respectively, $\funct(\timefreq)$ (cf. Eq.~\eqref{eq:U_g}), $\inflation(\timefreq)$ (Eq.~\eqref{eq:coherent_inflation}), $\quadratureangle(\timefreq)$ (Eq.~\eqref{eq:quadrature_angle}), $\betas_n(\timefreq)$ (Eq.~\eqref{eq:betas}), and $\sigmablurring(\timefreq) \freqtrap/\dissrateflucu$ both for the upper bounded (Eq.~\eqref{eq:Sigma_b_flucu}) and the fluctuating decoherence (Eq.~\eqref{eq:blurring_distance}) with $\dissrateloc=0$ and $25 \psd_1 = 2 \psd_2 d^2$ such that both terms in Eq.~\eqref{eq:Gamma_flucu} contribute equally to $\dissrateflucu$. The polygons in panel (a) indicate relevant instances of time, namely times when $\funct(\timefreq)$ becomes zero ($\timefreq = \tauthree$), times when $\inflation(\timefreq)$ is maximum ($\timefreq = \taud$) and the time halfway through the classical trajectory ($\timefreq = \taumax$).
    These instances of time appear as grid lines in the rest of the plots.}
    \label{fig:relevant_functions}
\end{figure}

To transition to the classical centroid frame, we need to evaluate the classical trajectory $\psveccl(\timefreq) = (\poscl(\timefreq),\momcl(\timefreq))^\transpose$ with $\psveccl(0) = \firstmoms(0)$. This classical trajectory can be solved analytically \cite{Hsu1960,Brizard2009} using the Jacobi elliptic functions~\cite{Whittaker1996}.
In Fig.~\ref{fig:relevant_functions}a, we display the classical trajectory, which is periodic and has an {\em avocado} shape. The analytical solution allows us to obtain analytical expressions for the timescales governing the classical dynamics, which are solely functions of $\pos_\textrm{s}/\dwmin$. We will be particularly interested in the time scale
\begin{align}
\taumax = \log \left(\frac{4\sqrt{2} \dwmin}{\pos_\textrm{s}}\right) + O[(\pos_\textrm{s}/d)^2],
\end{align}
which corresponds to the turning point of the classical trajectory of the centroid. It also corresponds to the half-period of the evolution, and the time at which the distance from the origin reaches its maximum, given by $\poscl(\taumax) = \sqrt{2\dwmin^2 - \pos_\textrm{s}^2}$.

In the classical centroid frame, the effective potential from Eq.~(\ref{eq:U_eff}) becomes time-dependent. The harmonic part of this time-dependent effective potential, given by Eq.~(\ref{eq:U_g}), has a dimensionless spring constant:
\begin{equation}
\funct(\timefreq) = -1 + 3 \left(\frac{ \poscl(\timefreq)}{\dwmin} \right)^2,
\end{equation}
which we illustrate in Fig.~\ref{fig:relevant_functions}(b). Initially, this effective harmonic potential is inverted ($\funct(\timefreq)<0$) until the time $\tauthree$, which is defined by $\poscl(\tauthree) = d/\sqrt{3}$. It then transforms into a harmonic potential ($\funct(\timefreq)>0$) until time $ 2 \taumax -\tauthree$, where it reverts to being inverted. This sequence of inverted, harmonic, and inverted potentials echoes the dynamics leveraged in the loop protocol proposed in~\cite{Weiss2021}.

Using the effective Harmonic potential, we can transition to the Gaussian frame. In this frame, we can compute both the delocalization length $\inflation(\timefreq)$ and angle $\quadratureangle(\timefreq)$, which are depicted in Fig.\ref{fig:relevant_functions}(c) and Fig.\ref{fig:relevant_functions}(d), respectively. The state initially expands (i.e., squeezes) until it reaches a maximum at $\timefreq = \taud$, which for large $\freqtrap/\freq$ is approximately given by
\begin{align}
\inflation_\star = \frac{1}{\sqrt{2}} \frac{\freqtrap}{\freq}\frac{\pos_\textrm{s}}{\dwmin}.\label{eq:eta_max}
\end{align}
The state then compresses (i.e., anti-squeezes) until the turning point $\taumax$. From here, the dynamics \old{repeats }\new{is mirrored} until $2\taumax$. Regarding the angle $\quadratureangle(\timefreq)$, we observe the anticipated behavior discussed in Sec.~\ref{subsec:validity}, i.e., it remains approximately constant (after the initial variation) until the first turning point, at which point it undergoes an approximated $\pi$-shift.

In the Gaussian frame, the strength of the nonlinear dynamics induced by the nonharmonicities in the potential is parameterized by the $\betas_n(\timefreq)$ ($n>2$) variables, as given in Eq.~(\ref{eq:betas}). For the double-well potential, we have
\begin{align}
\betas_3(\timefreq) &= \frac{\freq}{ \freqtrap \dwmin^2} 3 \poscl(\timefreq)\inflation(\timefreq)^{3},\label{eq:betathree}\\
\betas_4(\timefreq) &= \frac{\freq}{\freqtrap \dwmin^2} \inflation(\timefreq)^{4}.\label{eq:betafour}
\end{align}
These variables are illustrated in Fig.\ref{fig:relevant_functions}(e). They demonstrate how the strength of these nonlinearities is determined by the amount of squeezing $\inflation(\timefreq)$, which can be compared with Fig.\ref{fig:relevant_functions}(c). Furthermore, it is evident that the cubic nonharmonicity is larger than the quartic nonharmonicity in the small-fluctuations regime, denoted by $\inflation(\timefreq) \ll \poscl(\timefreq) $.

Regarding the impact of decoherence, the key variable within the linearized-decoherence approximation, which is valid within the regime $ \inflation(\timefreq) \ll \poscl(\timefreq)$, is the blurring distance $\sigmablurring(\timefreq)^2$, as shown in Eq.~\eqref{eq:blurring_distance}. The blurring distance depends on the effective time-dependent decoherence rate $\dissrateeff(\timefreq) = \dissrateloc+ \dissratefluc(\timefreq)$, which includes the contribution from the potential's fluctuations as given in Eq.~\eqref{eq:Gamma_fluc}. For the case of the double-well potential where $\poscl(\timefreq) < \sqrt{2}\dwmin$, $\dissratefluc(\timefreq)$ is upper bounded by the time-independent decoherence rate
\begin{equation} \label{eq:Gamma_flucu}
\frac{\dissratefluc(\timefreq)}{\freqtrap} < \frac{\dissrateflucu}{\freqtrap} \equiv \frac{\pi\freq^4 }{2\freqtrap^3} \left( 25 \psd_1 +2 \psd_2 \dwmin^2 \right).
\end{equation}
This allows us to conveniently bound the time-dependent decoherence rate by a time-independent rate, namely $\dissrateeff(\timefreq) < \dissrate \equiv \dissrateloc+ \dissrateflucu$, leading to an upper-bound blurring distance given by
\begin{align} \label{eq:Sigma_b_flucu}
\sigmablurring(\timefreq)^2 < \frac{4 \dissrate}{\freq} \int_0^\timefreq d\timefreq' \inflation(\timefreq')^2 .
\end{align}
In Fig.\ref{fig:relevant_functions}(f), we plot both the original blurring distance (dashed line) and the upper bound (solid line), showing that in the relevant time scale $\timefreq \approx \taumax$ the upper bound is a factor of $10$ larger than the original blurring distance.

To implement the constant-angle approximation, we need to set two angles: one, $\constantangle$, for $\timefreq \in [0,\taumax]$, and a second one, $\constantangle_2$, for $\timefreq \in (\taumax, 2 \taumax]$. As discussed in Sec.~\ref{subsec:validity}, and observed in Fig.\ref{fig:relevant_functions}d, $\constantangle - \constantangle_2 = \pi - \delta$, where $0< \delta \ll 1$ necessitates fine tuning. We choose $\constantangle = \quadratureangle(\taud)$; that is, at the time when $d{\quadratureangle}(\timefreq)/d \timefreq$ is at its first minimum (recall Eq.~\eqref{eq:angle_derivative} and that $\eta(\timefreq)$ is at its maximum at $\timefreq = \taud$), and we select $\delta$ so that the state at $\timefreq = 2 \taumax$ is closest to the one numerically calculated. For the parameters given in Table~\ref{tab:parameters}, this corresponds to $\constantangle/\pi= \quadratureangle(\taud)/\pi \approx 0.499969$ and $\delta/\pi = 7.63944 \times 10^{-6}$.

We are now in a position to analyze how well our analytical approach describes the numerically exact nonlinear dynamics of this particular example.

\subsection{Comparison of Analytical Approach with Numerically Exact Results}

\begin{figure}
    \centering
    \includegraphics[width=\linewidth]{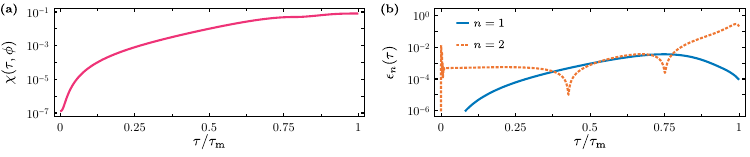}
    \caption{Quantities related to the validity of the constant-angle approximation for dynamics in a double-well with the parameters in Table~\ref{tab:parameters}. Panel (a) shows $\chi(\timefreq;\constantangle)$, and panel (b) shows the relative errors $\epsilon_1(\timefreq)$ (see Eq.~\eqref{eq:epsilon_1}) and $\epsilon_2(\timefreq)$ (see Eq.~\eqref{eq:epsilon_2}).}
    \label{fig:figure_of_merit}
\end{figure}

Following the discussion in Sec.~\ref{subsec:validity}, we first display in Fig.~\ref{fig:figure_of_merit}(a) the function $\chi(\timefreq; \constantangle)$, as defined in Eq.~\eqref{eq:chi}, evaluated for $\timefreq \in [0,\taumax]$. The fact that this quantity is small is an indicator of the validity of the constant-angle approximation. More importantly, in Fig.~\ref{fig:figure_of_merit}(b), we demonstrate the relative errors $\epsilon_n(\timefreq)$ for the first ($n=1$) and second ($n=2$) moments, as defined in Eq.~\eqref{eq:epsilon_1} and Eq.~\eqref{eq:epsilon_2}, for the case of dynamics without decoherence, which is the most sensitive case. The numerical calculations, performed using the split-operator method~\cite{Leforestier1991}, reveal relative errors well below one percent, with the exception of the second moments around the turning point $\timefreq=\taumax$. This is consistent with the fact that the constant-angle approximation is less accurate around a turning point. This is also evidenced in Fig.~\ref{fig:figure_of_merit}(a), where the increase of $\chi(\timefreq; \constantangle)$ towards the turning point is observable. These results suggest that our analytical approach should yield an excellent approximation of nonlinear open dynamics.

\begin{figure*}
    \centering
    \includegraphics[width=\linewidth]{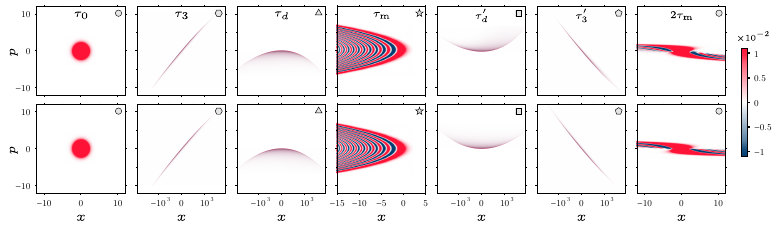}
    \caption{Wigner function of the state of a particle evolving in a double-well potential with parameters given in Table~\ref{tab:parameters} at different instances of time.
    These instances of time are indicated by polygons and correspond to the times indicated in Fig.~\ref{fig:relevant_functions}(a).
    The first row shows the numerically exact Wigner function $\wigner(\psvec + \psveccl(\timefreq),\timefreq)$ obtained using a numerically exact method whereas the second row shows the approximated Wigner function $\wigner_\nongaussian(\psvec + \psveccl(\timefreq),\timefreq)$ obtained using our analytical approach.}
    \label{fig:wigner_evolution}
\end{figure*}

To demonstrate this explicitly, we plot the Wigner function in the centroid frame at the six specific instances of time indicated in Fig.~\ref{fig:relevant_functions}a, as shown in Fig.~\ref{fig:wigner_evolution}. We present both the numerically exact results $\wigner(\psvec + \psveccl(\timefreq),\timefreq)$ (which require several hours of computation) and the results using our analytical approach $\wigner_\nongaussian(\psvec + \psveccl(\timefreq),\timefreq)$ (which only require a few minutes). The agreement between the two is remarkably excellent. Furthermore, the cubic-phase states~\cite{Weedbrook2012, Brunelli2019, Kala2022, Neumeier2024, Rakhubovsky2021}, generated during the times $\timefreq \in [0, \taumax]$, can be obtained analytically. Indeed, one can show, as detailed in App.~\ref{subapp:wigner_cubic}, that the state in both the centroid and Gaussian frames, in the absence of decoherence during this time scale, can be expressed as follows:
\begin{align}
\wignerharapprox(\psvec,\timefreq) = \frac{1}{\sqrt{2\pi}|\kappa_{3}(\timefreq)|^{1/3}} \exp\left[\frac{6\kappathree(\timefreq) \mom_\constantangle+ 1}{12|\kappa_{3}(\timefreq)|^2}\right] \Aifunc \left( \frac{\kappathree(\timefreq) \pos_\constantangle^2+\mom_\constantangle}{|\kappa_{3}(\timefreq)|^{1/3}} + \frac{1}{4|\kappa_{3}(\timefreq)|^{4/3}}\right).\label{eq:airy_pure}
\end{align}
Here, $\Aifunc(z)$ denotes the Airy function~\cite{Vallee2004}. For simplification, we have omitted the contribution of $\kappafour(\timefreq)$ in Eq.~\eqref{eq:airy_pure}, which is negligible for $\timefreq \leq \taumax$. The complete expression can be found in App.~\ref{subapp:wigner_cubic}. Equation~\eqref{eq:airy_pure} indeed represents the Wigner function of a cubic-phase state~\cite{Weedbrook2012,Brunelli2019,Kala2022,Neumeier2024,Rakhubovsky2021}. Interestingly, the state at $\timefreq = 2 \taumax$ exhibits a pronounced quartic character. This might seem surprising given that, as we mentioned before, $\kappathree(\timefreq) \gg \kappafour(\timefreq)$. However, upon inspecting the map $\mapng (\timefreq)$, one realizes that the two contributions of $\kappathree(\timefreq)$ and $\kappafour(\timefreq)$, corresponding to the first half-orbit ($0<\timefreq \leq \taumax$) and the second half-orbit ($\taumax< \timefreq \leq 2\taumax$), would exactly cancel out or add up for a change in the angle of precisely $\pi$ (i.e., $\delta = 0$). When $\delta$ is small but nonzero, the contribution from $\kappafour(\timefreq)$ still accumulates, while the difference of the cubic terms, to first order, contributes another quarticlike term, thereby giving the state at $2\taumax$ a quartic character.

\begin{figure}
    \centering
    \includegraphics[width=.6\linewidth]{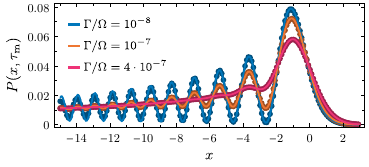}
    \caption{Position probability distribution at time $\taumax$ for a state evolving in a double-well potential for the parameters in Table~\ref{tab:parameters} and for different values of $\dissrate$. The lines are computed using the analytical method described in this paper, whereas dots correspond to a numerically exact computation using Q-Xpanse~\cite{RodaLlordes2023_2}.}
    \label{fig:px}
\end{figure}

Finally, an interesting feature of the cubic-phase state generated at $\timefreq = \taumax$ is that it produces an interference pattern in the position probability distribution $\psprob(\pos,\timefreq) = \int d\mom \wigner(\psvec,\timefreq)$. This is depicted in Fig.~\ref{fig:px} for different values of $\dissrate/\freqtrap$, using both the numerically exact Q-Xpanse method~\cite{RodaLlordes2023_2} (dots) and our analytical approach (solid line). This demonstrates not only excellent qualitative agreement but also quantitative, particularly for the first and most relevant interference fringes. With our analytical approach, it can be shown that the interference pattern in the presence of decoherence, $\psprob_\approximated(\pos,\timefreq) = \int d\mom \wigner_\approximated(\psvec,\timefreq)$, can be expressed as a Gaussian convolution of the interference pattern in the absence of decoherence~\cite{Romero-Isart2011}, $\psprob_{\coh,\approximated}(\pos,\timefreq)$, as follows:
\begin{align}
\psprob_\approximated(\pos,\timefreq) = \frac{1}{\sqrt{2\pi C_{\textit{b},\pos\pos}(\timefreq)}}\int d\pos' \exp\left[-\frac{(\pos-\pos')^2}{2C_{\textit{b},\pos\pos}(\timefreq)}\right] \psprob_{\coh,\approximated}(\pos',\timefreq).\label{eq:psprob_convolution}
\end{align}
Here, $C_{\textit{b},\pos\pos}(\timefreq)$ is the $xx$ component of the blurring covariance matrix given in Eq.~\eqref{eq:noise_matrix}, and
\begin{equation}
\psprob_{\coh,\approximated}(\pos,\timefreq) \equiv \int d\mom \mapcl(\timefreq)\maphar(\timefreq) \mapng (\timefreq) \wignerhar(\psvec,0),
\end{equation}
corresponds to the probability distribution in the absence of decoherence, which is explicitly provided in App.~\ref{subapp:prob_cubic}. In addition, the interference pattern in the absence of decoherence can be derived from Eq.~\eqref{eq:airy_pure} and is given by:
\begin{align}
\psprob_{\coh,\approximated}(\pos,\timefreq) \propto \exp\left(-\dfrac{\pos-\poscl(\timefreq)}{2\mxp(\timefreq)\kappas_3(\timefreq)}\right) \left|\Aifunc\left(\frac{\pos-\poscl(\timefreq)}{\mxp(\timefreq)(4\kappas_3(\timefreq))^{1/3}} + z(\timefreq)\right)\right|^2.\label{eq:pos_prob}
\end{align}
In this equation, $\mxp(\timefreq) = \propagatorcomp_{\pos\pos}(\timefreq) \sin(\constantangle) - \propagatorcomp_{xp}(\timefreq) \cos(\constantangle)$, and $z(\timefreq) \in \mathbb{C}$ is a time-dependent function that is given in App.~\ref{subapp:prob_cubic}. This analytical expression allows us to derive the scaling of the fringe separation, $\fringesep$, which is the distance between the largest interference peak in $\psprob_{\coh,\approximated}(\pos,\taumax)$ and its second one. As shown in App.~\ref{subapp:prob_cubic}, the fringe separation scales as $\fringesep \propto |\kappathree(\timefreq)|^{1/3}$, or in terms of the physical parameters as $\fringesep \propto (\freqtrap/\freq)^{2/3}(1/\dwmin)^{1/3}$, for a fixed $\pos_\textrm{s}/\dwmin$.

\section{Conclusions} \label{sec:conclusions}

We have provided an analytical treatment of the Wigner function dynamics for a continuous-variable degree of freedom, such as the position and momentum of a massive particle, in a nonharmonic potential and in the presence of decoherence caused by, among other things, fluctuations of the nonharmonic potential. Our analytical treatment has demonstrated that it can yield an expression of the time-evolved Wigner function that approximates the exact dynamics very accurately, particularly in the case of wide nonharmonic potentials and small fluctuations.

Our analytical method specifically entails performing an exact, suitable reformulation of the open nonlinear dynamical problem via two frame transformations: the classical centroid and the Gaussian frame transformations. In this transformed frame, two approximations can be applied: the key constant-angle approximation and the linearized-decoherence approximation. These approximations facilitate the integration of the dynamical problem and are argued to provide excellent approximations for dynamics that allow the state to expand over scales many orders of magnitude larger than the initial spatial expansion, provided these fluctuations are smaller than the available phase space that can be explored. In other words, the fluctuations of the state should not occupy the entire available phase-space surface. These requirements define the concept of wide nonharmonic potentials and small fluctuations.

We have tested our analytical method using an example of a massive particle evolving in a wide double-well potential. The method shows excellent agreement with numerical simulations, not just qualitatively, but also quantitatively. In the wide regime, especially in the presence of decoherence, the numerical calculations can be challenging. For these calculations, we utilized a numerical tool, Q-Xpanse, that we recently developed~\cite{RodaLlordes2023_2}. These numerical calculations often require several hours of computation, while our analytical approach reproduces the results with simple evaluations requiring only a few minutes of calculation. This example is particularly relevant as it has been recently proposed for preparing macroscopic quantum states of a levitated nanoparticle via the generated nonlinear quantum dynamics~\cite{RodaLlordes2023}. The combination of the numerical tool and the analytical method has enabled us to design, optimize, and understand this experimental proposal.

Beyond its utility in modeling experiments, our analytical method also paves the way for interesting research questions that warrant further investigation. Specifically, it would be intriguing to better characterize the dynamical problems in which our method proves effective. It could potentially be utilized to define a class of semiclassical quantum dynamical problems (see e.g., \cite{Heller1975, Berry1977}). Furthermore, one could concentrate on the classical regime, either by taking the limit as $\hbar \rightarrow 0$ and studying the efficacy of this method in understanding classical nonlinear dynamics, or by augmenting the strength of the generators of decoherence. In the latter case, it would also be interesting to focus on different types of decoherence, including non-Gaussian decoherence, and apply our method to understand and characterize their effect on the dynamics. 

In conclusion, we hope that our analytical approach to understanding open quantum nonlinear dynamics and its classical limit will significantly contribute to this captivating, albeit long-standing, topic.

\begin{acknowledgments}
ARC and MRL contributed equally to this work. We thank the Q-Xtreme synergy group for fruitful discussions. This research was supported by the European Union’s Horizon 2020 research and innovation programme under grant agreement No. [863132] (IQLev) and by the European Research Council (ERC) under the grant Agreement No. [951234] (Q-Xtreme ERC-2020-SyG). ARC acknowledges funding from the European Union’s Horizon 2020 research and innovation programme under the Marie Skłodowska-Curie grant agreement No. [101103589] (DecoXtreme, HORIZON-MSCA-2022-PF-01-01). PTG was partially supported by the Foundation for Polish Science (FNP).
\end{acknowledgments}

\appendix

\section{Explicit expression of the decoherence rates} \label{app:dissipator}

In the main body of the text, we have reformulated the dissipator due to the fluctuation potential (cf. Eq.~\eqref{eq:decoh_dissip_expansion}) as a double sum with coefficients denoted by $\dissrate_{nm}$. To achieve this, we performed a Taylor expansion of the potential, $\potential(\posunits)$, as well as its derivative in Eq.~\eqref{eq:dissipatior_fluctuations}. This led us to the explicit form of the equation:
\begin{equation}
\dissrate_{nm} = \frac{2 \pi}{\hbar^2} \frac{\zppos^{m+n}}{m!n!} \left[ A_1 \zppos^2 \frac{\partial^{m+1} \potential}{\partial \posunits^{m+1}}(0)\frac{\partial^{n+1} \potential}{\partial \posunits^{n+1}}(0) + A_2 \frac{\partial^{m} \potential}{\partial \posunits^m}(0)\frac{\partial^{n} \potential}{\partial \posunits^n}(0) \right] + \delta_{n,1}\delta_{m,1} \dissrateloc.
\end{equation}
These coefficients $\dissrate_{nm}$ are particularly convenient for expressing the decoherence superoperator within the Hilbert space. However, their translation to the Wigner equation is considerably different, necessitating the introduction of a new set of coefficients denoted as $\coeffnmk$ in Eq.~\eqref{eq:dissipator_wigner_space}. These can be derived by applying the Wigner transform to the decoherence superoperator $\dissipator[\cdot]$, resulting in the following equation:
\begin{equation}
\coeffnmk = (-i)^{k}\left[1+(-)^{k}\right] \frac{\dissrate_{nm}}{2\freq} \left[ \sum_{q=0}^{k} (-)^{q} \binom{n}{k} \binom{m}{k-q} - \binom{m+n}{k} \right].
\end{equation}

\section{Details on the constant-angle approximation }\label{app:constant_angle_approximation}

Here, we discuss in detail the constant-angle approximation. The dynamics in a wide potential is such that the coherent expansion $\inflation(\timefreq) \gg 1$, and therefore the angle $\quadratureangle(\timefreq)$ is approximately constant except close to compression times. Hence, one can approximate $\quadratureangle(\timefreq)$ as a piecewise time-dependent function. Let $\taumax$ be the time at which the only turning point between $0$ and $\timefreq_\mathrm{f}$ occurs. Then, we first use the \textit{exact} factorization property of the evolution map 
\begin{align}
    \map_+(\timefreq_\mathrm{f}) = \exp_+\left[g\int_0^{\timefreq_\mathrm{f}} \lin(\timefreq_1) d\timefreq_1\right] = \exp_+\left[g\int_{\taumax}^{\timefreq_\mathrm{f}} \lin(\timefreq_1) d\timefreq_1\right]\exp_+\left[g\int_0^{\taumax} \lin (\timefreq_1) d\timefreq_1\right],
\end{align}
for a generic $g\lin(\timefreq)$. Now in each time interval, $[0,\taumax]$ and $[\taumax,\timefreq_\mathrm{f}]$ the angle is approximately constant. Namely, our analytic calculation is based on the replacement of $M_+(\timefreq_\mathrm{f})$ by 
\begin{align}
     \map(\timefreq_\mathrm{f}) = \exp\left[\int_{\timefreq_m}^{\timefreq_\mathrm{f}} \mathcal{L}(\timefreq_1) d\timefreq_1\right]\exp\left[\int_0^{\timefreq_m} \mathcal{L}(\timefreq_1) d\timefreq_1\right],
\end{align}
together with the approximation
\begin{align}
    \varphi(\timefreq) \approx \begin{cases}
        \phi_1 & \text{for } \timefreq \in [0,\timefreq_m),\\
        \phi_2 & \text{for } \timefreq > \timefreq_m
    \end{cases}.
\end{align}
for two constant angles $\phi_1$ and $\phi_2$. Note that we have reduced the problem to the scenario in which there are no compression times in the integration region. In what follows, let us assume that there are no compression times from between $0$ and $\timefreq$. 

The constant angle approximation described in the main text has the effect of replacing a time-ordered exponential, which is difficult to calculate, by an easier simple exponential map. Mathematically, the reason why this is possible can be justified as follows. Consider the expansion of the time-ordered exponential map
\begin{align}
    \map_+(\timefreq) = \toexp\left[g \int_0^\timefreq d\timefreq_1 \lin(\timefreq_1) \right] = \sum_{n=0}^\infty g^n \int_0^\timefreq d\timefreq_1 \cdots \int_0^{\timefreq_{n-1}} d\timefreq_n \lin(\timefreq_1) \cdots \lin(\timefreq_n),
\end{align}
with the time-dependent generator $g \lin(\timefreq)$. The corresponding ``standard'' exponential map is instead
\begin{align}
    \map(\timefreq) = \exp\left[g \int_0^\timefreq d\timefreq_1 \lin(\timefreq_1)  \right] = \sum_{n=0}^\infty \frac{g^n}{n!} \int_0^\timefreq d\timefreq_1 \cdots \int_0^{\timefreq} d\timefreq_n \lin(\timefreq_1) \cdots \lin(\timefreq_n).
\end{align}
Note that if $\lin(\timefreq)$ commutes with itself at different times, both expressions coincide after redefining the integration region. In the general case, the difference between the two maps is given by
\begin{align}
    \map_+(\timefreq) - \map(\timefreq) = \frac{g^2}{2}\int_0^\timefreq d\timefreq_1 \int_0^{\timefreq_1} d\timefreq_2 [\lin(\timefreq_1), \lin(\timefreq_2)] + O(g^3)\label{eq:map_commutator},
\end{align}
which explicitly shows that the first correction depends on the out-of-time commutator of the generator. 

In our case of interest $\lin(\timefreq)= \linharnG(\timefreq) + \linhard(\timefreq)$, where $\lin(\timefreq)$ includes only multiplication with respect to $\pos_{\quadratureangle(\timefreq)}$ and (multiple) derivatives with respect to $\mom_{\quadratureangle(\timefreq)}$. Hence, the term
\begin{align}
    \frac{\partial}{\partial \mom_{\quadratureangle(\timefreq_1)}} \pos(\timefreq_2) = \sin(\quadratureangle(\timefreq_1)-\quadratureangle(\timefreq_2)),
\end{align}
controls the strength of the commutator. The explicit calculation is lengthy and therefore we do not include it here. However, let us show how to proceed by considering only the ``classical'' term defined as
\begin{align}
    \lin^{(\Gaussian)}_\classical(\timefreq) \equiv \sum_{n\geq2}\betas_{n+1}(\timefreq) \pos_{\quadratureangle(\timefreq)}^n \frac{\partial}{\partial \mom_{\quadratureangle(\timefreq)}},
\end{align}
since the rest are analogous. We find that the first-order correction in Eq.~\eqref{eq:map_commutator} is given by
\begin{align}
    \frac{1}{2}\int_0^\timefreq d\timefreq_1 \int_0^{\timefreq_1} d\timefreq_2 [\lin^{(\Gaussian)}_\classical(\timefreq_1), \lin^{(\Gaussian)}_\classical(\timefreq_2)] = \frac{1}{2}\int_0^\timefreq d\timefreq_1 \int_0^{\timefreq_1} d\timefreq_2 A(\timefreq_1,\timefreq_2) \sin(\quadratureangle(\timefreq_1)-\quadratureangle(\timefreq_2)),
\end{align}
where we have defined the differential operator
\begin{align}
    A(\timefreq_1,\timefreq_2) = \sum_{n,m\geq 2} \betas_{n+1}(\timefreq_1) \betas_{m+1}(\timefreq_2)\pos_{\quadratureangle(\timefreq_1)}^{n-1} \pos_{\quadratureangle(\timefreq_2)}^{m-1} \left[m \pos_{\quadratureangle(\timefreq_1)} \frac{\partial}{\partial \mom_{\quadratureangle(\timefreq_2)}} + n \pos_{\quadratureangle(\timefreq_2)} \frac{\partial}{\partial \mom_{\quadratureangle(\timefreq_1)}}\right].
\end{align}
In the regime of large expansions $\eta(\timefreq)\gg 1$ allowed by wide potentials, the angle $\quadratureangle(\timefreq)$ is approximately constant  (see Eq.~\eqref{eq:angle_derivative}). Hence, replacing the time-dependent angle $\quadratureangle(\timefreq)$ by a constant $\constantangle$ corresponds to replacing the time-ordered exponential by a standard exponential map, which in turn allows to continue with the analytical calculation.

The approximation still depends on the value of the angle $\constantangle$ that one chooses. We decide to take $\constantangle$ as the angle $\quadratureangle(\timefreq_\star)$ evaluated at the time of largest expansion $\timefreq_\star = \mathrm{arg}\!\max_\timefreq \eta(\timefreq)$, since this corresponds to the time at which the derivative of $\quadratureangle(\timefreq)$ is the smallest.

\section{An alternative linearized-decoherence approximation requirement}\label{app:underlying_reason}

The small fluctuations condition $\poscl(\timefreq)\gg\inflation(\timefreq)$ ensures that decoherence differential operator $\linhard(\timefreq)$ can be linearized (see Eq.~\eqref{eq:linearized_decoherence_gaussian}). However, a drawback of this condition is that $\poscl(\timefreq)$ is defined relative to an arbitrary origin. While the coefficients $c_{nmk}$ (and consequently $\Gamma_{nm}$) appropriately account for this arbitrariness, it can potentially obscure the more general condition for the linearized-decoherence approximation to hold. Here, we provide a more intricate derivation of the operator $\intlind(\timefreq)$ in Eq.~\eqref{eq:noise_map_general}, but that has the advantage of eliminating the arbitrariness associated with the choice of origin.

Let $f(\posunitsop/\zppos)$ be a smooth function of the position operator $\posunitsop$, and let us compute the Wigner transform
\begin{align}
    f(\posunitsop/\zppos) \state f(\posunitsop/\zppos) \mapsto  f\left(\pos + i 
\frac{\partial}{\partial \mom}\right) f\left(\pos -i 
\frac{\partial}{\partial \mom}\right) \wigner(\psvec). 
\end{align}
Using this relation, a double commutator term reads
\begin{align}
    - [f(\posunitsop),[f(\posunitsop), \cdot]] \mapsto \Xi\cdot \equiv \sum_{s,s'=\pm} (-ss') f\left(\pos + i s 
\frac{\partial}{\partial \mom}\right)f\left(\pos + i  s'
\frac{\partial}{\partial \mom}\right) \cdot.
\end{align}
In the dissipator due to fast fluctuating forces, we have two such terms: $f(\pos) \propto (\partial \varpotential/\partial \pos)(\pos)$ and $f(\pos) \propto \varpotential(\pos)$. The transformation to the classical centroid and Gaussian frames amounts to transforming the expression $\Xi$ into
\begin{align}
    \Xi^\gaussianframe(\timefreq) = \sum_{s,s'=\pm 1} (-ss') f\bigl(\poscl(\timefreq) + \eta(\timefreq) Z_s(\quadratureangle(\timefreq)) \bigr)  f\bigl(\poscl(\timefreq) + \inflation(\timefreq) Z_{s'}(\inflation(\timefreq))\bigr) \cdot.\label{eq:dissipator_formal}
\end{align}
with the short-hand notation
\begin{align}
    Z_s(\constantangle) = \left(\pos_{\phi} + i s
\frac{\partial}{\partial p_{\constantangle}}\right).
\end{align}
One can now expand Eq.~\eqref{eq:dissipator_formal} around $\poscl(\timefreq)$ to obtain 
\begin{align}
    \Xi^\gaussianframe(\timefreq) = \sum_{n,m=0}^\infty \left(\frac{\partial^n f}{\partial \pos^n}(\poscl(\timefreq)) \inflation(\timefreq)^n\right)\left(\frac{\partial^m f}{\partial \pos^m}(\poscl(\timefreq)) \inflation(\timefreq)^m\right) \sum_{ss'=\pm}(-ss')Z_s(\phi(t))^n Z_{s'}(\phi(t))^m .\label{eq:decoherence_series}
\end{align}
If we label the terms in the series as $(n,m)$, the first nonvanishing term is $(1,1)$. The linearized decoherence approximation corresponds to keeping the term $(1,1)$ and assuming that the rest are subdominant. In essence, we require that the terms of the sequence $\{a_n\}$ with 
\begin{align}
    a_n = \inflation(\timefreq)^n \frac{\partial^n f}{\partial \pos^n}(\poscl(\timefreq)),
\end{align}
decreases for increasing $n$, where $f(\pos)$ is proportional to either the potential $\varpotential(\pos)$ or its derivative. This is expected in the limit of wide potentials. Indeed, let $d$ be the length-scale of the potential, then we have
\begin{align}
    \inflation(\timefreq)^{n} \frac{\partial^{n} f}{\partial \pos^{n}}(\poscl(t)) \sim \left(\frac{\inflation(\timefreq)}{\dwmin}\right)^n f(\poscl(\timefreq)).
\end{align}
Therefore, the term with the least number of factors $\inflation(\timefreq)/\dwmin$ is the term $(1,1)$, and we expect it to dominate the series. Note that the above logic does not exploit that $\poscl(\timefreq) \gg \inflation(\timefreq)$, but rather that $\inflation(\timefreq)$ is small as compared to the derivatives of the potential $\varpotential(\pos)$; that is, that the potential is wide.

\section{Derivations using the analytical expression of the Wigner function}

This appendix is devoted to providing additional details regarding the derivations of analytical expressions for the approximated first and second moments, along with the approximated Wigner function, using the constant-angle approximation; that is, using our main result in Eq.~\eqref{eq:W_a}. In the derivation, we use the properties
\begin{align}
    &\int d\psvec f(\psvec) \{\gaussian[\covariance](\psvec) \star g(\psvec)]\} = \int d\psvec \{\gaussian[\covariance](\psvec) \star f(\psvec)\} g(\psvec), \label{eq:convolution_property} \\
    &\int d\psvec f(\psvec) g(T^{-1}\psvec) = \int d\psvec f(T\psvec) g(\psvec) J_T(\psvec) \label{eq:transformation_property},
\end{align}
where $\star$ denotes the convolution,
\begin{align}
    f(\psvec)\star g(\psvec) =\int d\psvec' f(\psvec-\psvec') g(\psvec'),
\end{align}
and the Jacobian $J_T(\psvec)$ is the determinant of the matrix 
\begin{align}
\frac{\partial(T\psvec)}{\partial(\psvec)} = \begin{pmatrix}
        \partial_\pos (T\psvec)_\pos & \partial_\mom (T\psvec)_\pos\\
        \partial_\pos (T\psvec)_\mom & \partial_\mom (T\psvec)_\mom
    \end{pmatrix}
\end{align}
associated to the transformation $T$.

\subsection{First moments}\label{subapp:first_moms}

In the Wigner representation, expectation values of symmetrically-ordered (also Weyl-ordered) operators are computed as phase space integrals. Using our main result in Eq.~\eqref{eq:W_a}, we find that
\begin{align}
    \firstmoms_a(\timefreq) \equiv \average{\hat{\psvec}}_a(\timefreq) = \int d\psvec \psvec \new{\mapnoiseapprox}(\timefreq) \mapcl(\timefreq)\maphar(\timefreq) \mapng (\timefreq) \wignerhar(\psvec,0).
\end{align}
Using the property in Eq.~\eqref{eq:convolution_property} together with the convolution representation of the decoherence map in Eq.~\eqref{eq:convolution}, one can show that the decoherence map does not contribute to the evaluation of the first moments. Moreover, using the property in Eq.~\eqref{eq:transformation_property} for the maps $\mapcl(\timefreq)$ and $\maphar(\timefreq)$, and noting that both transformations have unit Jacobian, one arrives to
\begin{align}
    \firstmoms_a(\timefreq) = \psveccl(\timefreq) + \propagator(\timefreq) \int d\psvec \mapng(\timefreq) \wignerhar(\psvec,0).
\end{align}
Finally, we note that, in our approximation, the quantum terms included in $\mapng(\timefreq)$ do not contribute to the calculation of the first or second moments. This is a consequence of the fact that the quantum terms include derivatives of order higher than three, which after integration by parts would give no contribution to the integral. Hence, the action of the map $\mapng(\timefreq)$ amounts to a nonlinear transformation 
\begin{align}
    T\psvec = \psvec - \rotation(\phi)^\transpose \etwo \sum_{n\geq 3} \kappas_n(\timefreq) (\eone^\transpose \rotation(\phi) \psvec)^{n-1}\label{eq:transformation_explicit}
\end{align}
which also has a unit Jacobian. Hence, combining our previous results and taking the average with the state $\wigner_g(\psvec,0)$ we arrive at the result quoted in Eq.~\eqref{eq:first_moments_nG}.

\subsection{Covariance matrix}\label{subapp:second_moms}

The covariance matrix is computed as the average
\begin{align}
    \covariance_a(\timefreq) \equiv& \average{(\hat{\psvec}-\firstmoms(\timefreq))(\hat{\psvec}-\firstmoms(\timefreq))^\transpose}(\timefreq)\nonumber\\
    =& \int d\psvec (\psvec-\firstmoms(\timefreq))(\psvec-\firstmoms(\timefreq))^\transpose \new{\mapnoiseapprox}(\timefreq) \mapcl(\timefreq)\maphar(\timefreq) \mapng (\timefreq) \wignerhar(\psvec,0).
\end{align}
Taking advantage of the property in Eq.~\eqref{eq:convolution_property}, we see that the action of the decoherence map is to increase the value of the covariance matrix by $\Sigmablurring(\timefreq)$, as compared to the case without decoherence. Namely, one arrives at 
\begin{align}
    \covariance_a(\timefreq) = \Sigmablurring(\timefreq) + \int d\psvec (\psvec-\firstmoms(\timefreq))(\psvec-\firstmoms(\timefreq))^\transpose  \mapcl(\timefreq)\maphar(\timefreq) \mapng (\timefreq) \wignerhar(\psvec,0).
\end{align}
Moreover, similarly to the case of the first moments, taking advantage of Eq.~\eqref{eq:transformation_property} and Eq.~\eqref{eq:transformation_explicit}, one can manipulate the result into the form displayed in Eq.~\eqref{eq:second_moments_nG}.

\subsection{State before the turning point in the double-well case}\label{subapp:wigner_cubic}

Here, we give more details on how to obtain an analytical expression for the Wigner function in the centroid and Gaussian frame under the constant-angle approximation. To this end, we assume a thermal initial state $\wigner_g(\psvec,0)$, which we decompose into Fourier modes as
\begin{align}
    \wignerhar(\psvec,0) = \frac{\exp[-\pos_\phi^2/(2(2\nth+1))]}{\sqrt{2\pi(2\nth+1)}} \int \frac{d\conjmom_\phi}{2\pi} \exp\left[-(2\nth + 1) \conjmom_\phi^2/2 + i \mom_\phi \conjmom_\phi \right],\label{eq:initial_wigner_fourier}
\end{align}
Applying the evolution map given by the exponential of $\intlin(\timefreq)$ in Eq.~\eqref{eq:map_approx_ng} to the state in Eq.~\eqref{eq:initial_wigner_fourier} leads to an integral expression of the form
\begin{align}
    \frac{1}{2\pi} \int dk \exp\left[i\left( c_3 \frac{k^3}{3} + i c_2 \frac{k^2}{2} + c_1 k\right)\right] = \frac{1}{|c_3|^{1/3}} \exp\left(\frac{c_2^3 + 6 c_3 c_2 c_1}{12 |c_3|^2}\right) \Aifunc\left(\frac{c_2^2+4 c_3 c_1}{4|c_3|^{4/3}}\right),\label{eq:Gauss_airy_integral}
\end{align}
where $c_3 \equiv \kappathree(\timefreq) + 3\kappas_4(\timefreq) \pos_\phi$, $c_2 \equiv 2\nth + 1 + \sigmablurring(\timefreq)^2$, and $c_1 \equiv \mom_\phi + \kappathree(\timefreq) \pos_\phi^2+\kappas_4(\timefreq) \pos_\phi^3$ and we have introduced the Airy function
\begin{align}
    \Aifunc(z) \equiv \frac{1}{2\pi} \int_{-\infty}^\infty du \exp\left[i\left(\frac{u^3}{3} + zu\right)\right],\label{eq:airy_def}
\end{align}
for $z\in\mathbb{C}$. Multiplying by the remaining Gaussian in $\pos_\phi$, one can simplify the expression into the final result
\begin{align}
    \wigner^\gaussianframe_\approximated(\psvec,\timefreq) =& \frac{1}{\sqrt{2\pi(2\nth+1)} |\kappa_{3}(\timefreq)+3\kappas_4(\timefreq)\pos_\phi|^{1/3}} \nonumber\\
    &\times \exp\left[\frac{(2\nth+1 + \sigmablurring(\timefreq)^2)[6(\kappathree(\timefreq)+3\kappas_4(\timefreq)\pos_\phi) (\mom_\phi + \kappathree(\timefreq) \pos_\phi^2+\kappas_4(\timefreq) \pos_\phi^3)]}{12|\kappathree(\timefreq)+3\kappas_4(\timefreq)\pos_\phi|^2}\right]\nonumber\\
    &\times \exp\left[\frac{(2\nth+1 + \sigmablurring(\timefreq)^2)^3}{12|\kappathree(\timefreq)+3\kappas_4(\timefreq)\pos_\phi|^2}\right] \exp\left[-\frac{\pos_\phi^2}{2\standev^2}\right] \nonumber\\
    &\times \Aifunc \left( \frac{(\kappathree(\timefreq) \pos_\phi^2+\kappafour(\timefreq)\pos_\phi^3+\mom_\phi)}{|\kappathree(\timefreq)+3\kappas_4(\timefreq)\pos_\phi|^{1/3}} +  \frac{(2\nth+1 + \sigmablurring(\timefreq)^2)^2}{4|\kappathree(\timefreq)+3\kappas_4(\timefreq)\pos_\phi|^{4/3}}\right).\label{eq:wigner_ip_final}
\end{align}
If one sets into the above expression $\kappafour(\timefreq) \mapsto 0$, $\sigmablurring(\timefreq)\mapsto0$, and $\nth \mapsto 0$ it yields the simplified result
\begin{align}
    \wigner^\gaussianframe_\approximated(\psvec,\timefreq) =& \frac{1}{\sqrt{2\pi} |\kappa_{3}(\timefreq)|^{1/3}} \exp\left[\frac{6\kappathree(\timefreq) \mom_\constantangle+ 1}{12|\kappathree(\timefreq)|^2}\right]  \Aifunc \left( \frac{\kappathree(\timefreq) \pos_\constantangle^2+\mom_\constantangle}{|\kappathree(\timefreq)|^{1/3}} +  \frac{1}{4|\kappathree(\timefreq)|^{4/3}}\right),\label{eq:airy_pure_app}
\end{align}
which we show in the main text. Finally, one can move back to the centroid frame by applying $\maphar(\timefreq)$, and to the original frame by applying $\mapcl(\timefreq)$. It follows that the approximated Wigner function yields
\begin{align}
    \wigner_\approximated(\psvec,\timefreq) = \wigner^\gaussianframe_\approximated(\propagator(\timefreq)^{-1}(\psvec-\psveccl(\timefreq)),\timefreq).\label{eq:wigner_cubic_phase_state}
\end{align}

\subsection{Interference pattern at the turning point in the double-well case}\label{subapp:prob_cubic}

In this Appendix, we compute the coherent interference pattern of the probability distribution $\psprob_{\coh,\approximated}(\pos,\timefreq)$ within our approximation. To obtain the desired probability distribution, we first rewrite $\pos_\phi$ and $\mom_\phi$ in terms of $\pos$ and $\mom$ using the rotation matrix $\rotation(\phi)$. Then, we apply the map $\maphar(\timefreq)$ to move to the classical trajectory frame, which mixes the coordinates $\pos$ and $\mom$ according to the propagator $\propagator(\timefreq)^{-1}$. If we denote $f(\psvec_\phi)$ the function of the right-hand-side of Eq.~\eqref{eq:airy_pure_app}, the Wigner function in the classical centroid frame is $f(\textbf{m}(\timefreq)\psvec)$, where $\textbf{m}(\timefreq) \equiv \rotation(\phi) \propagator(\timefreq)^{-1}$, with components $m_{ij}(\timefreq)$ for $i,j = \pos,\mom$. For conciseness of the notation, hereafter we denote $m_{ij}(\timefreq)$ simply as $m_{ij}$.

In order to obtain the probability distribution, we have to perform the integral over the variable $\mom$, which can be done using the following property of the Airy function~(see Ch.3 of \cite{Vallee2004})
\begin{align}
    \int dy \Aifunc(y^2 + y_0) \exp(i k y) = 2^{2/3} \pi \Aifunc[2^{-2/3}(y_0 + k)]\Aifunc[2^{-2/3}(y_0 - k)].\label{eq:airy_y2}
\end{align}
In order to use the property in Eq.~\eqref{eq:airy_y2}, we have to rewrite the argument of the Airy function in Eq.~\eqref{eq:airy_pure} without a linear term. To this end, we define 
\begin{align}
    &y = \frac{1}{\kappathree(\timefreq)^{1/6}}\left(\kappathree(\timefreq)^{1/2} \mxp \mom + \frac{2 \kappathree(\timefreq) \mxp \mxx \pos + \mpp}{2 \mxp \kappathree(\timefreq)^{1/2}} \right),\\
    &y_0(\pos) = \frac{1}{\kappathree(\timefreq)^{1/3}}\left(\frac{\mxp^2- \mpp^2-4 \mxp \kappathree(\timefreq) \pos }{4 \mxp^2 \kappathree(\timefreq)}\right),\\
    &k = -i \frac{\mpp}{2 \mxp \kappathree(\timefreq)^{4/3}}.
\end{align}
Noting that $\mxx \mpp - \mxp \mpx = 1$, and using $y$ as an integration variable, we can use formula Eq.~\eqref{eq:airy_y2} to compute
\begin{align}
    \text{P}^\centroidframe_{\coh,\approximated}(\pos,\timefreq) =& \frac{\sqrt{\pi/2}}{|\mxp|(\kappathree(\timefreq)/2)^{2/3}} \exp\left[\frac{\mxp^2-3 \mpp^2-6\mxp \kappathree(\timefreq) \pos}{12 \mxp^2 \kappathree(\timefreq)^2}\right] \nonumber\\
    &\times \left|\Aifunc\left[\frac{1}{(4\kappathree(\timefreq))^{1/3}}\left(-\frac{\pos}{\mxp} + \frac{1}{4\kappathree(\timefreq)}\left(1-\frac{\mpp^2}{\mxx^2}\right)+i  \frac{\mpp}{2 \mxp \kappathree(\timefreq)}\right)\right]\right|^2.\label{eq:position_distribution}
\end{align}
Introducing back the expressions for $\mxx$, $\mxp$, $\mpx$ and $\mpp$, and shifting $\pos \mapsto \pos-\poscl(\timefreq)$ yields the probability $\text{P}_{\coh,\approximated}(\pos,\timefreq)$. The effect of decoherence can be included with Eq.~\eqref{eq:psprob_convolution}. 

Let $\fringesep$ be the distance between the first and second maximum in the fringe pattern defined by Eq.~\eqref{eq:position_distribution}. At $\taumax$, the dependence on the position variable $\pos$ is scaled with 
\begin{align}
    \fringesep \sim |\kappathree(\taumax)|^{1/3} \mxp.
\end{align}
with $\mxp \approx S_{\pos\pos}(\taumax)$, which is only a function of $\pos_\textrm{s}/\dwmin$. Moreover, one has that
\begin{align}
    \kappathree(\taumax) = 3 \frac{\freq}{\freqtrap} \frac{1}{\dwmin} \int_0^{\taumax} d\timefreq' \eta(\timefreq')^3 \frac{\poscl(\timefreq')}{\dwmin} = 3 \frac{\freq}{\freqtrap} \frac{1}{\dwmin} \eta_\star^3 \int_0^{\taumax} d\timefreq' \frac{\eta(\timefreq')^3}{\eta_\star^3} \frac{\poscl(\timefreq')}{\dwmin}.\label{eq:kappa3_at_tmax}
\end{align}
For sufficiently large $\freqtrap/\freq$ the integral in Eq.~\eqref{eq:kappa3_at_tmax} becomes, in very good approximation, also a function of only $\pos_\textrm{s}/\dwmin$. In that regime, we find that the fringe separation scales with
\begin{align}
    \fringesep \sim \left( \frac{\freqtrap}{\freq}\right)^{2/3} \left(\frac{1}{d}\right)^{1/3} f(\pos_\textrm{s}/\dwmin),
\end{align}
where $f(\pos_\textrm{s}/\dwmin)$ is an unspecified function of only $\pos_\textrm{s}/\dwmin$.

\printbibliography

@article{Wigner1932,
  title = {On the Quantum Correction For Thermodynamic Equilibrium},
  author = {Wigner, E.},
  journal = {Phys. Rev.},
  volume = {40},
  issue = {5},
  pages = {749--759},
  numpages = {0},
  year = {1932},
  publisher = {American Physical Society},
  doi = {10.1103/PhysRev.40.749},
  url = {https://link.aps.org/doi/10.1103/PhysRev.40.749}
}

@book{Schleich2001,
author = {Schleich,  Wolfgang P.},
publisher = {John Wiley \& Sons, Ltd},
isbn = {9783527602971},
title = {Quantum Optics in Phase Space},
doi = {https://doi.org/10.1002/3527602976},
url = {https://onlinelibrary.wiley.com/doi/abs/10.1002/3527602976},
year = {2001},
}

@article{Case2008,
    author = {Case, William B.},
    title = "{Wigner functions and Weyl transforms for pedestrians}",
    journal = {Am. J. Phys.},
    volume = {76},
    number = {10},
    pages = {937-946},
    year = {2008},
    month = {10},
    issn = {0002-9505},
    doi = {10.1119/1.2957889},
    url = {https://doi.org/10.1119/1.2957889}
}

@inbook{Zurek2007,
	address = {Basel},
	author = {Zurek, Wojciech Hubert},
	booktitle = {Quantum Decoherence: Poincar{\'e} Seminar 2005},
	doi = {10.1007/978-3-7643-7808-0_1},
	isbn = {978-3-7643-7808-0},
	pages = {1--31},
	publisher = {Birkh{\"a}user Basel},
	title = {Decoherence and the Transition from Quantum to Classical --- Revisited},
	url = {https://doi.org/10.1007/978-3-7643-7808-0_1},
	year = {2007},
    %editor = {Duplantier, Bertrand and Raimond, Jean-Michel and Rivasseau, Vincent},
}

@article{Zurek1988,
  title = {Decoherence, Chaos, and the Correspondence Principle},
  author = {Habib, Salman and Shizume, Kosuke and Zurek, Wojciech Hubert},
  journal = {Phys. Rev. Lett.},
  volume = {80},
  issue = {20},
  pages = {4361--4365},
  numpages = {0},
  year = {1998},
  publisher = {American Physical Society},
  doi = {10.1103/PhysRevLett.80.4361},
  url = {https://link.aps.org/doi/10.1103/PhysRevLett.80.4361}
}

@article{Berry1977,
author = {Berry, Michael Victor},
title = {Semi-classical mechanics in phase space: A study of Wigner’s function},
journal = {Philos. Trans. Royal Soc. A},
volume = {287},
number = {1343},
pages = {237-271},
year = {1977},
doi = {10.1098/rsta.1977.0145},
URL = {https://royalsocietypublishing.org/doi/abs/10.1098/rsta.1977.0145}
}

@article{Schrodinger1926,
  title = {An Undulatory Theory of the Mechanics of Atoms and Molecules},
  author = {Schr\"odinger, E.},
  journal = {Phys. Rev.},
  volume = {28},
  issue = {6},
  pages = {1049--1070},
  numpages = {0},
  year = {1926},
  publisher = {American Physical Society},
  doi = {10.1103/PhysRev.28.1049},
  url = {https://link.aps.org/doi/10.1103/PhysRev.28.1049}
}

@article{Heisenberg1927,
	author = {Heisenberg, W. },
	da = {1927/03/01},
	doi = {10.1007/BF01397280},
	id = {Heisenberg1927},
	journal = {Z. Phys.},
	number = {3},
	pages = {172--198},
	title = {{\"U}ber den anschaulichen Inhalt der quantentheoretischen Kinematik und Mechanik},
	ty = {JOUR},
	url = {https://doi.org/10.1007/BF01397280},
	volume = {43},
	year = {1927}}

@article{Born1926,
	author = {Born, Max},
	da = {1926/11/01},
	doi = {10.1007/BF01397184},
	id = {Born1926},
	journal = {Z. Phys.},
	number = {11},
	pages = {803--827},
	title = {Quantenmechanik der Sto{\ss}vorg{\"a}nge},
	ty = {JOUR},
	url = {https://doi.org/10.1007/BF01397184},
	volume = {38},
	year = {1926}}

@article{Unruh1989,
  title = {Reduction of a wave packet in quantum Brownian motion},
  author = {Unruh, W. G. and Zurek, W. H.},
  journal = {Phys. Rev. D},
  volume = {40},
  issue = {4},
  pages = {1071--1094},
  numpages = {0},
  year = {1989},
  publisher = {American Physical Society},
  doi = {10.1103/PhysRevD.40.1071},
  url = {https://link.aps.org/doi/10.1103/PhysRevD.40.1071}
}

@article{RodaLlordes2023,
  title = {Macroscopic Quantum Superpositions via Dynamics in a Wide Double-Well Potential},
  author = {Roda-Llordes, M. and Riera-Campeny, A. and Candoli, D. and Grochowski, P. T. and Romero-Isart, O.},
  journal = {Phys. Rev. Lett.},
  volume = {132},
  issue = {2},
  pages = {023601},
  numpages = {7},
  year = {2024},
  month = {Jan},
  publisher = {American Physical Society},
  doi = {10.1103/PhysRevLett.132.023601},
  url = {https://link.aps.org/doi/10.1103/PhysRevLett.132.023601}
}

@article{RodaLlordes2023_2,
  title = {Numerical simulation of large-scale nonlinear open quantum mechanics},
  author = {Roda-Llordes, M. and Candoli, D. and Grochowski, P. T. and Riera-Campeny, A. and Agrenius, T. and Garc\'{\i}a-Ripoll, J. J. and Gonzalez-Ballestero, C. and Romero-Isart, O.},
  journal = {Phys. Rev. Res.},
  volume = {6},
  issue = {1},
  pages = {013262},
  numpages = {10},
  year = {2024},
  month = {Mar},
  publisher = {American Physical Society},
  doi = {10.1103/PhysRevResearch.6.013262},
  url = {https://link.aps.org/doi/10.1103/PhysRevResearch.6.013262}
}

@article{Schneider1999,
  title = {Decoherence and fidelity in ion traps with fluctuating trap parameters},
  author = {Schneider, S. and Milburn, G. J.},
  journal = {Phys. Rev. A},
  volume = {59},
  issue = {5},
  pages = {3766--3774},
  numpages = {0},
  year = {1999},
  publisher = {American Physical Society},
  doi = {10.1103/PhysRevA.59.3766},
  url = {https://link.aps.org/doi/10.1103/PhysRevA.59.3766}
}

@article{Henkel1999,
	author = {Henkel, C. and P{\"o}tting, S. and Wilkens, M.},
	da = {1999/12/01},
	doi = {10.1007/s003400050823},
	id = {Henkel1999},
	journal = {App. Phys. B},
	number = {5},
	pages = {379--387},
	title = {Loss and heating of particles in small and noisy traps},
	ty = {JOUR},
	url = {https://doi.org/10.1007/s003400050823},
	volume = {69},
	year = {1999}}

@article{Deleglise2008,
	author = {Del{\'e}glise, Samuel and Dotsenko, Igor and Sayrin, Cl{\'e}ment and Bernu, Julien and Brune, Michel and Raimond, Jean-Michel and Haroche, Serge},
	da = {2008/09/01},
	doi = {10.1038/nature07288},
	id = {Del{\'e}glise2008},
	journal = {Nature},
	number = {7212},
	pages = {510--514},
	title = {Reconstruction of non-classical cavity field states with snapshots of their decoherence},
	ty = {JOUR},
	url = {https://doi.org/10.1038/nature07288},
	volume = {455},
	year = {2008}}

@article{Hofheinz2009,
	author = {Hofheinz, Max and Wang, H. and Ansmann, M. and Bialczak, Radoslaw C. and Lucero, Erik and Neeley, M. and O'Connell, A. D. and Sank, D. and Wenner, J. and Martinis, John M. and Cleland, A. N.},
	da = {2009/05/01},
	doi = {10.1038/nature08005},
	id = {Hofheinz2009},
	journal = {Nature},
	number = {7246},
	pages = {546--549},
	title = {Synthesizing arbitrary quantum states in a superconducting resonator},
	ty = {JOUR},
	url = {https://doi.org/10.1038/nature08005},
	volume = {459},
	year = {2009}}

@article{Kirchmair2013,
	author = {Kirchmair, Gerhard and Vlastakis, Brian and Leghtas, Zaki and Nigg, Simon E. and Paik, Hanhee and Ginossar, Eran and Mirrahimi, Mazyar and Frunzio, Luigi and Girvin, S. M. and Schoelkopf, R. J.},
	da = {2013/03/01},
	doi = {10.1038/nature11902},
	id = {Kirchmair2013},
	journal = {Nature},
	number = {7440},
	pages = {205--209},
	title = {Observation of quantum state collapse and revival due to the single-photon Kerr effect},
	ty = {JOUR},
	url = {https://doi.org/10.1038/nature11902},
	volume = {495},
	year = {2013}}

@article{Leibfied1996,
  title = {Experimental Determination of the Motional Quantum State of a Trapped Atom},
  author = {Leibfried, D. and Meekhof, D. M. and King, B. E. and Monroe, C. and Itano, W. M. and Wineland, D. J.},
  journal = {Phys. Rev. Lett.},
  volume = {77},
  issue = {21},
  pages = {4281--4285},
  numpages = {0},
  year = {1996},
  publisher = {American Physical Society},
  doi = {10.1103/PhysRevLett.77.4281},
  url = {https://link.aps.org/doi/10.1103/PhysRevLett.77.4281}
}

@article{Fluhmann2020,
  title = {Direct Characteristic-Function Tomography of Quantum States of the Trapped-Ion Motional Oscillator},
  author = {Fl\"uhmann, C. and Home, J. P.},
  journal = {Phys. Rev. Lett.},
  volume = {125},
  issue = {4},
  pages = {043602},
  numpages = {6},
  year = {2020},
  publisher = {American Physical Society},
  doi = {10.1103/PhysRevLett.125.043602},
  url = {https://link.aps.org/doi/10.1103/PhysRevLett.125.043602}
}

@article{Heller1975,
    author = {Heller, Eric J.},
    title = {Time‐dependent approach to semiclassical dynamics},
    journal = {J. Chem. Phys.},
    volume = {62},
    number = {4},
    pages = {1544-1555},
    year = {1975},
    month = {02},
    issn = {0021-9606},
    doi = {10.1063/1.430620},
    url = {https://doi.org/10.1063/1.430620}
}

@article{Heller1976,
    author = {Heller, Eric J.},
    title = {Wigner phase space method: Analysis for semiclassical applications},
    journal = {J. Chem. Phys.},
    volume = {65},
    number = {4},
    pages = {1289-1298},
    year = {1976},
    month = {08},
    issn = {0021-9606},
    doi = {10.1063/1.433238},
    url = {https://doi.org/10.1063/1.433238}
}

@article{Dania2023,
  title = {Ultrahigh Quality Factor of a Levitated Nanomechanical Oscillator},
  author = {Dania, Lorenzo and Bykov, Dmitry S. and Goschin, Florian and Teller, Markus and Kassid, Abderrahmane and Northup, Tracy E.},
  journal = {Phys. Rev. Lett.},
  volume = {132},
  issue = {13},
  pages = {133602},
  numpages = {7},
  year = {2024},
  month = {Mar},
  publisher = {American Physical Society},
  doi = {10.1103/PhysRevLett.132.133602},
  url = {https://link.aps.org/doi/10.1103/PhysRevLett.132.133602}
}

@article{Kamba2023,
	author = {Kamba, Mitsuyoshi and Shimizu, Ryoga and Aikawa, Kiyotaka},
	da = {2023/12/01},
	doi = {10.1038/s41467-023-43745-7},
	id = {Kamba2023},
	isbn = {2041-1723},
	journal = {Nat. Commun.},
	number = {1},
	pages = {7943},
	title = {Nanoscale feedback control of six degrees of freedom of a near-sphere},
	ty = {JOUR},
	url = {https://doi.org/10.1038/s41467-023-43745-7},
	volume = {14},
	year = {2023}}

@article{Hu1992,
  title = {Quantum Brownian motion in a general environment: Exact master equation with nonlocal dissipation and colored noise},
  author = {Hu, B. L. and Paz, Juan Pablo and Zhang, Yuhong},
  journal = {Phys. Rev. D},
  volume = {45},
  issue = {8},
  pages = {2843--2861},
  numpages = {0},
  year = {1992},
  publisher = {American Physical Society},
  doi = {10.1103/PhysRevD.45.2843},
  url = {https://link.aps.org/doi/10.1103/PhysRevD.45.2843}
}

@article{vanKampen1976,
	author = {N.G {Van Kampen}},
	doi = {https://doi.org/10.1016/0370-1573(76)90029-6},
	issn = {0370-1573},
	journal = {Phys. Rep.},
	number = {3},
	pages = {171-228},
	title = {Stochastic differential equations},
	url = {https://www.sciencedirect.com/science/article/pii/0370157376900296},
	volume = {24},
	year = {1976}}

@article{Maurer2022,
  title = {Quantum theory of light interaction with a Lorenz-Mie particle: Optical detection and three-dimensional ground-state cooling},
  author = {Maurer, Patrick and Gonzalez-Ballestero, Carlos and Romero-Isart, Oriol},
  journal = {Phys. Rev. A},
  volume = {108},
  issue = {3},
  pages = {033714},
  numpages = {26},
  year = {2023},
  month = {Sep},
  publisher = {American Physical Society},
  doi = {10.1103/PhysRevA.108.033714},
  url = {https://link.aps.org/doi/10.1103/PhysRevA.108.033714}
}

@article{Agrenius2023,
  title = {Interaction between an Optically Levitated Nanoparticle and Its Thermal Image: Internal Thermometry via Displacement Sensing},
  author = {Agrenius, Thomas and Gonzalez-Ballestero, Carlos and Maurer, Patrick and Romero-Isart, Oriol},
  journal = {Phys. Rev. Lett.},
  volume = {130},
  issue = {9},
  pages = {093601},
  numpages = {8},
  year = {2023},
  publisher = {American Physical Society},
  doi = {10.1103/PhysRevLett.130.093601},
  url = {https://link.aps.org/doi/10.1103/PhysRevLett.130.093601}
}

@article{Weiss2019,
  title = {Quantum motional state tomography with nonquadratic potentials and neural networks},
  author = {Weiss, Talitha and Romero-Isart, Oriol},
  journal = {Phys. Rev. Res.},
  volume = {1},
  issue = {3},
  pages = {033157},
  numpages = {10},
  year = {2019},
  publisher = {American Physical Society},
  doi = {10.1103/PhysRevResearch.1.033157},
  url = {https://link.aps.org/doi/10.1103/PhysRevResearch.1.033157}
}

@article{Kala2022,
	author = {Vojt\v{e}ch Kala and Radim Filip and Petr Marek},
	doi = {10.1364/OE.464759},
	journal = {Opt. Express},
	number = {17},
	pages = {31456--31471},
	publisher = {Optica Publishing Group},
	title = {Cubic nonlinear squeezing and its decoherence},
	url = {https://opg.optica.org/oe/abstract.cfm?URI=oe-30-17-31456},
	volume = {30},
	year = {2022}}

@article{Moore2022,
	author = {Moore, Darren W. and Filip, Radim},
	da = {2022/05/30},
	doi = {10.1038/s42005-022-00910-6},
	id = {Moore2022},
	journal = {Comm. Phys.},
	number = {1},
	pages = {128},
	title = {Hierarchy of quantum non-Gaussian conservative motion},
	ty = {JOUR},
	url = {https://doi.org/10.1038/s42005-022-00910-6},
	volume = {5},
	year = {2022}}

@article{RomeroIsart2011,
  title = {Quantum superposition of massive objects and collapse models},
  author = {Romero-Isart, Oriol},
  journal = {Phys. Rev. A},
  volume = {84},
  issue = {5},
  pages = {052121},
  numpages = {17},
  year = {2011},
  publisher = {American Physical Society},
  doi = {10.1103/PhysRevA.84.052121},
  url = {https://link.aps.org/doi/10.1103/PhysRevA.84.052121}
}

@article{Delic2020,
    author = {Uroš Delić  and Manuel Reisenbauer  and Kahan Dare  and David Grass  and Vladan Vuletić  and Nikolai Kiesel  and Markus Aspelmeyer},
    title = {Cooling of a levitated nanoparticle to the motional quantum ground state},
    journal = {Science},
    volume = {367},
    number = {6480},
    pages = {892-895},
    year = {2020},
    doi = {10.1126/science.aba3993},
    URL = {https://www.science.org/doi/abs/10.1126/science.aba3993}}

@article{Kamba2022,
    author = {Mitsuyoshi Kamba and Ryoga Shimizu and Kiyotaka Aikawa},
    journal = {Opt. Express},
    number = {15},
    pages = {26716--26727},
    publisher = {Optica Publishing Group},
    title = {Optical cold damping of neutral nanoparticles near the ground state in an optical lattice},
    volume = {30},
    year = {2022},
    url = {https://opg.optica.org/oe/abstract.cfm?URI=oe-30-15-26716},
    doi = {10.1364/OE.462921}
}

@article{Ranfagni2022,
  title = {Two-dimensional quantum motion of a levitated nanosphere},
  author = {Ranfagni, A. and B\o{}rkje, K. and Marino, F. and Marin, F.},
  journal = {Phys. Rev. Res.},
  volume = {4},
  issue = {3},
  pages = {033051},
  numpages = {10},
  year = {2022},
  publisher = {American Physical Society},
  doi = {10.1103/PhysRevResearch.4.033051},
  url = {https://link.aps.org/doi/10.1103/PhysRevResearch.4.033051}
}

@article{Piotrowski2023,
	author = {Piotrowski, Johannes and Windey, Dominik and Vijayan, Jayadev and Gonzalez-Ballestero, Carlos and de los R{\' i}os Sommer, Andr{\'e}s and Meyer, Nadine and Quidant, Romain and Romero-Isart, Oriol and Reimann, Ren{\'e} and Novotny, Lukas},
	da = {2023/07/01},
	doi = {10.1038/s41567-023-01956-1},
	id = {Piotrowski2023},
	journal = {Nat. Phys.},
	number = {7},
	pages = {1009--1013},
	title = {Simultaneous ground-state cooling of two mechanical modes of a levitated nanoparticle},
	ty = {JOUR},
	url = {https://doi.org/10.1038/s41567-023-01956-1},
	volume = {19},
	year = {2023}}

@article{vanKampen1974,
title = {A cumulant expansion for stochastic linear differential equations. II},
journal = {Physica},
volume = {74},
number = {2},
pages = {239-247},
year = {1974},
issn = {0031-8914},
doi = {https://doi.org/10.1016/0031-8914(74)90122-0},
url = {https://www.sciencedirect.com/science/article/pii/0031891474901220},
author = {N.G. {Van Kampen}}
}

@article{Tebbenjohanns2021,
  title = {Quantum Control of a Nanoparticle Optically Levitated in Cryogenic Free Space},
  author = {Tebbenjohanns, Felix and Mattana, M. Luisa and Rossi, Massimiliano and Frimmer, Martin and Novotny, Lukas},
  year = {2021},
  month = jul,
  journal = {Nature},
  volume = {595},
  number = {7867},
  pages = {378--382},
  publisher = {{Nature Publishing Group}},
  issn = {1476-4687},
  doi = {10.1038/s41586-021-03617-w}
  }

@article{Magrini2021,
	author = {Magrini, Lorenzo and Rosenzweig, Philipp and Bach, Constanze and Deutschmann-Olek, Andreas and Hofer, Sebastian G. and Hong, Sungkun and Kiesel, Nikolai and Kugi, Andreas and Aspelmeyer, Markus},
	da = {2021/07/01},
	doi = {10.1038/s41586-021-03602-3},
	id = {Magrini2021},
	journal = {Nature},
	number = {7867},
	pages = {373--377},
	title = {Real-time optimal quantum control of mechanical motion at room temperature},
	ty = {JOUR},
	url = {https://doi.org/10.1038/s41586-021-03602-3},
	volume = {595},
	year = {2021}}

@article{Gonzalez-Ballestero2021,
author = {C. Gonzalez-Ballestero  and M. Aspelmeyer  and L. Novotny  and R. Quidant  and O. Romero-Isart },
title = {Levitodynamics: Levitation and control of microscopic objects in vacuum},
journal = {Science},
volume = {374},
number = {6564},
pages = {eabg3027},
year = {2021},
doi = {10.1126/science.abg3027},
URL = {https://www.science.org/doi/abs/10.1126/science.abg3027}}

@article{Zurek2001,
	author = {Zurek, Wojciech Hubert},
	da = {2001/08/01},
	doi = {10.1038/35089017},
	id = {Zurek2001},
	journal = {Nature},
	number = {6848},
	pages = {712--717},
	title = {Sub-Planck structure in phase space and its relevance for quantum decoherence},
	url = {https://doi.org/10.1038/35089017},
	volume = {412},
	year = {2001}}

@article{Brunelli2019,
  title = {Linear and quadratic reservoir engineering of non-Gaussian states},
  author = {Brunelli, Matteo and Houhou, Oussama},
  journal = {Phys. Rev. A},
  volume = {100},
  issue = {1},
  pages = {013831},
  numpages = {17},
  year = {2019},
  publisher = {American Physical Society},
  doi = {10.1103/PhysRevA.100.013831},
  url = {https://link.aps.org/doi/10.1103/PhysRevA.100.013831}
}

@article{Weedbrook2012,
  title = {Gaussian quantum information},
  author = {Weedbrook, Christian and Pirandola, Stefano and Garc\'{i}a-Patr\'on, Ra\'ul and Cerf, Nicolas J. and Ralph, Timothy C. and Shapiro, Jeffrey H. and Lloyd, Seth},
  journal = {Rev. Mod. Phys.},
  volume = {84},
  issue = {2},
  pages = {621--669},
  numpages = {0},
  year = {2012},
  publisher = {American Physical Society},
  doi = {10.1103/RevModPhys.84.621},
  url = {https://link.aps.org/doi/10.1103/RevModPhys.84.621}
}

@article{Brizard2009,
doi = {10.1088/0143-0807/30/4/007},
url = {https://dx.doi.org/10.1088/0143-0807/30/4/007},
year = {2009},
publisher = {},
volume = {30},
number = {4},
pages = {729},
author = {Alain J Brizard},
title = {A primer on elliptic functions with applications in classical mechanics},
journal = {Eur. J. Phys.}
}

@article{Hsu1960,
 ISSN = {0033569X, 15524485},
 URL = {http://www.jstor.org/stable/43634717},
 author = {C. S. Hsu},
 journal = {Q. Appl. Math.},
 number = {4},
 pages = {393--407},
 publisher = {Brown University},
 title = {ON THE APPLICATION OF ELLIPTIC FUNCTIONS IN NON-LINEAR FORCED OSCILLATIONS},
 urldate = {2023-02-24},
 volume = {17},
 year = {1960}
}

@article{Weiss2021,
  title = {Large Quantum Delocalization of a Levitated Nanoparticle Using Optimal Control: Applications for Force Sensing and Entangling via Weak Forces},
  author = {Weiss, T. and Roda-Llordes, M. and Torrontegui, E. and Aspelmeyer, M. and Romero-Isart, O.},
  journal = {Phys. Rev. Lett.},
  volume = {127},
  issue = {2},
  pages = {023601},
  numpages = {6},
  year = {2021},
  publisher = {American Physical Society},
  doi = {10.1103/PhysRevLett.127.023601},
  url = {https://link.aps.org/doi/10.1103/PhysRevLett.127.023601}
}

@article{Joos1985,
	author = {Joos, E. and Zeh, H. D.},
	doi = {10.1007/BF01725541},
	journal = {Z. Phys. B},
	number = {2},
	pages = {223--243},
	title = {The emergence of classical properties through interaction with the environment},
	ty = {JOUR},
	url = {https://doi.org/10.1007/BF01725541},
	volume = {59},
	year = {1985}}

@book{Schlosshauer2007,
    title = {Decoherence and the Quantum-To-Classical Transition},
    author = {M. Schlosshauer},
    year = {2007},
    publisher = {Springer-Verlag},
    address = {Berlin, Heidelberg},
    isbn = {978-3-540-35773-5},
    doi = {https://doi.org/10.1007/978-3-540-35775-9}
}

@article{Pino2018,
doi = {10.1088/2058-9565/aa9d15},
url = {https://dx.doi.org/10.1088/2058-9565/aa9d15},
year = {2018},
publisher = {IOP Publishing},
volume = {3},
number = {2},
pages = {025001},
author = {H Pino and J Prat-Camps and K Sinha and B Prasanna Venkatesh and O Romero-Isart},
title = {On-chip quantum interference of a superconducting microsphere},
journal = {Quantum Sci. Technol.}
}

@article{Neumeier2024,
author = {Lukas Neumeier  and Mario A. Ciampini  and Oriol Romero-Isart  and Markus Aspelmeyer  and Nikolai Kiesel },
title = {Fast quantum interference of a nanoparticle via optical potential control},
journal = {Proc. Natl. Acad. Sci. U.S.A.},
volume = {121},
number = {4},
pages = {e2306953121},
year = {2024},
doi = {10.1073/pnas.2306953121},
URL = {https://www.pnas.org/doi/abs/10.1073/pnas.2306953121},
eprint = {https://www.pnas.org/doi/pdf/10.1073/pnas.2306953121}}

@article{Jain2016,
  title = {Direct Measurement of Photon Recoil from a Levitated Nanoparticle},
  author = {Jain, Vijay and Gieseler, Jan and Moritz, Clemens and Dellago, Christoph and Quidant, Romain and Novotny, Lukas},
  journal = {Phys. Rev. Lett.},
  volume = {116},
  issue = {24},
  pages = {243601},
  numpages = {5},
  year = {2016},
  publisher = {American Physical Society},
  doi = {10.1103/PhysRevLett.116.243601},
  url = {https://link.aps.org/doi/10.1103/PhysRevLett.116.243601}
}

@article{Romero-Isart2017,
doi = {10.1088/1367-2630/aa99bf},
url = {https://dx.doi.org/10.1088/1367-2630/aa99bf},
year = {2017},
publisher = {IOP Publishing},
journal = {New J. Phys.},
volume = {19},
number = {12},
pages = {123029},
author = {Oriol Romero-Isart}
}

@article{Gehm1998,
  title = {Dynamics of noise-induced heating in atom traps},
  author = {Gehm, M. E. and O'Hara, K. M. and Savard, T. A. and Thomas, J. E.},
  journal = {Phys. Rev. A},
  volume = {58},
  issue = {5},
  pages = {3914--3921},
  numpages = {0},
  year = {1998},
  publisher = {American Physical Society},
  doi = {10.1103/PhysRevA.58.3914},
  url = {https://link.aps.org/doi/10.1103/PhysRevA.58.3914}
}

@article{Bateman2014,
  title = {Near-Field Interferometry of a Free-Falling Nanoparticle from a Point-like Source},
  author = {Bateman, James and Nimmrichter, Stefan and Hornberger, Klaus and Ulbricht, Hendrik},
  year = {2014},
  month = sep,
  journal = {Nat. Commun.},
  volume = {5},
  number = {1},
  pages = {4788},
  publisher = {{Nature Publishing Group}},
  issn = {2041-1723},
  doi = {10.1038/ncomms5788},
  copyright = {2014 Nature Publishing Group, a division of Macmillan Publishers Limited. All Rights Reserved.},
  langid = {english}
}

@article{Romero-Isart2011,
  title = {Large {{Quantum Superpositions}} and {{Interference}} of {{Massive Nanometer-Sized Objects}}},
  author = {{Romero-Isart}, O. and Pflanzer, A. C. and Blaser, F. and Kaltenbaek, R. and Kiesel, N. and Aspelmeyer, M. and Cirac, J. I.},
  year = {2011},
  journal = {Phys. Rev. Lett.},
  volume = {107},
  number = {2},
  pages = {020405},
  publisher = {{American Physical Society}},
  doi = {10.1103/PhysRevLett.107.020405}
}

@article{Leforestier1991,
  title = {A Comparison of Different Propagation Schemes for the Time Dependent {{Schr\"odinger}} Equation},
  author = {Leforestier, C and Bisseling, R.H and Cerjan, C and Feit, M.D and Friesner, R and Guldberg, A and Hammerich, A and Jolicard, G and Karrlein, W and Meyer, H.-D and Lipkin, N and Roncero, O and Kosloff, R},
  year = {1991},
  month = may,
  journal = {J. Comput. Phys.},
  volume = {94},
  number = {1},
  pages = {59--80},
  issn = {00219991},
  doi = {10.1016/0021-9991(91)90137-A},
  urldate = {2023-03-13}
}

@article{Caldeira1983,
    title = {Path integral approach to quantum Brownian motion},
    journal = {Physica A},
    volume = {121},
    number = {3},
    pages = {587-616},
    year = {1983},
    issn = {0378-4371},
    doi = {https://doi.org/10.1016/0378-4371(83)90013-4},
    url = {https://www.sciencedirect.com/science/article/pii/0378437183900134},
    author = {A.O. Caldeira and A.J. Leggett}
}

@book{Breuer2002,
  title={The Theory of Open Quantum Systems},
  author={Breuer, H.P. and Petruccione, F.},
  isbn={9780198520634},
  lccn={2002075713},
  doi = {10.1093/acprof:oso/9780199213900.001.0001},
  year={2002},
  publisher={Oxford University Press}
}

@book{Whittaker1996, 
place={Cambridge},
edition={4}, 
series={Cambridge Mathematical Library},
title={A Course of Modern Analysis}, 
DOI={10.1017/CBO9780511608759},
publisher={Cambridge University Press},
author={Whittaker, E. T. and Watson, G. N.},
year={1996}, 
collection={Cambridge Mathematical Library}}

@book{Vallee2004,
author = {Vallée, Olivier and Soares, Manuel},
title = {Airy Functions and Applications to Physics},
publisher={Imperial College Press},
year = {2010},
edition = {2},
place = {London},
doi = {10.1142/p709},
isbn={978-1-84816-548-9},
URL = {https://www.worldscientific.com/doi/abs/10.1142/p709}
}

@article{Rakhubovsky2021,
	author = {Rakhubovsky, Andrey A. and Filip, Radim},
	da = {2021/07/29},
	doi = {10.1038/s41534-021-00453-8},
	id = {Rakhubovsky2021},
	isbn = {2056-6387},
	journal = {npj Quantum Inf.},
	number = {1},
	pages = {120},
	title = {Stroboscopic high-order nonlinearity for quantum optomechanics},
	ty = {JOUR},
	url = {https://doi.org/10.1038/s41534-021-00453-8},
	volume = {7},
	year = {2021}}

@book{Mukamel1999,
    author = {S. Mukamel},
    title = {Principles of Nonlinear Optical Spectroscopy},
    publisher = {Oxford University Press},
    year = {1999},
    isbn = {978-0195132915},
    address = {New York}
}

@article{Keller1985,
 ISSN = {00361445},
 URL = {http://www.jstor.org/stable/2031056},
 author = {Joseph B. Keller},
 journal = {SIAM Review},
 number = {4},
 pages = {485--504},
 publisher = {Society for Industrial and Applied Mathematics},
 title = {Semiclassical Mechanics},
 urldate = {2024-01-29},
 volume = {27},
 year = {1985}
}

@book{Heller2018,
 ISBN = {9780691163734},
 URL = {http://www.jstor.org/stable/j.ctvc77gwd},
 author = {Eric J. Heller},
 publisher = {Princeton University Press},
 title = {The Semiclassical Way to Dynamics and Spectroscopy},
 urldate = {2024-01-29},
 year = {2018}
}

@book{Zworski2022,
  title={Semiclassical analysis},
  author={Zworski, Maciej},
  volume={138},
  year={2022},
  publisher={American Mathematical Society},
}

@article{Berry1972,
doi = {10.1088/0034-4885/35/1/306},
url = {https://dx.doi.org/10.1088/0034-4885/35/1/306},
year = {1972},
month = {jan},
publisher = {},
volume = {35},
number = {1},
pages = {315},
author = {M.V. Berry and K. E. Mount},
title = {Semiclassical approximations in wave mechanics},
journal = {Rep. Prog. Phys.},
}

@article{Heller1977,
    author = {Heller, Eric J.},
    title = "{Phase space interpretation of semiclassical theory}",
    journal = {J. Chem. Phys.},
    volume = {67},
    number = {7},
    pages = {3339-3351},
    year = {1977},
    month = {08},
    issn = {0021-9606},
    doi = {10.1063/1.435296},
    url = {https://doi.org/10.1063/1.435296},
    eprint = {https://pubs.aip.org/aip/jcp/article-pdf/67/7/3339/11096195/3339\_1\_online.pdf},
}

@article{Davis1984,
    author = {Davis, M. J. and Heller, E. J.},
    title = {Comparisons of classical and quantum dynamics for initially localized states},
    journal = {J. Chem. Phys.},
    volume = {80},
    number = {10},
    pages = {5036-5048},
    year = {1984},
    month = {05},
    issn = {0021-9606},
    doi = {10.1063/1.446571},
    url = {https://doi.org/10.1063/1.446571},
    eprint = {https://pubs.aip.org/aip/jcp/article-pdf/80/10/5036/11271378/5036\_1\_online.pdf},
}

@article{Huber1988,
    author = {Huber, Daniel and Heller, Eric J. and Littlejohn, Robert G.},
    title = {Generalized Gaussian wave packet dynamics, Schrödinger equation, and stationary phase approximation},
    journal = {J. Chem. Phys.},
    volume = {89},
    number = {4},
    pages = {2003-2014},
    year = {1988},
    month = {08},
    issn = {0021-9606},
    doi = {10.1063/1.455714},
    url = {https://doi.org/10.1063/1.455714},
    eprint = {https://pubs.aip.org/aip/jcp/article-pdf/89/4/2003/11257346/2003\_1\_online.pdf},
}

@article{Littlejohn1986,
	author = {R.G. Littlejohn},
	doi = {https://doi.org/10.1016/0370-1573(86)90103-1},
	issn = {0370-1573},
	journal = {Phys. Rep.},
	number = {4},
	pages = {193-291},
	title = {The semiclassical evolution of wave packets},
	url = {https://www.sciencedirect.com/science/article/pii/0370157386901031},
	volume = {138},
	year = {1986}}

@article{Gallis1990,
  title = {Environmental and spontaneous localization},
  author = {Gallis, Michael R. and Fleming, Gordon N.},
  journal = {Phys. Rev. A},
  volume = {42},
  issue = {1},
  pages = {38--48},
  numpages = {0},
  year = {1990},
  month = {Jul},
  publisher = {American Physical Society},
  doi = {10.1103/PhysRevA.42.38},
  url = {https://link.aps.org/doi/10.1103/PhysRevA.42.38}
}

@book{Maslov2001,
  title={Semi-classical approximation in quantum mechanics},
  author={Maslov, Victor P and Fedoriuk, Mikhail Vasilevich},
  volume={7},
  year={1981},
  publisher={D. Reidel publishing company}, 
  address = {London, England},
  isbn = {978-90-277-1219-6},
}

@article{Feynman1951,
  title = {An Operator Calculus Having Applications in Quantum Electrodynamics},
  author = {Feynman, Richard P.},
  journal = {Phys. Rev.},
  volume = {84},
  issue = {1},
  pages = {108--128},
  numpages = {0},
  year = {1951},
  month = {Oct},
  publisher = {American Physical Society},
  doi = {10.1103/PhysRev.84.108},
  url = {https://link.aps.org/doi/10.1103/PhysRev.84.108}
}

@article{vanKampen1974a,
	author = {N.G. {Van Kampen}},
	doi = {https://doi.org/10.1016/0031-8914(74)90121-9},
	issn = {0031-8914},
	journal = {Physica},
	number = {2},
	pages = {215-238},
	title = {A cumulant expansion for stochastic linear differential equations. I},
	url = {https://www.sciencedirect.com/science/article/pii/0031891474901219},
	volume = {74},
	year = {1974}}

@article{Cabrera2015,
  title = {Efficient method to generate time evolution of the Wigner function for open quantum systems},
  author = {Cabrera, Renan and Bondar, Denys I. and Jacobs, Kurt and Rabitz, Herschel A.},
  journal = {Phys. Rev. A},
  volume = {92},
  issue = {4},
  pages = {042122},
  numpages = {10},
  year = {2015},
  month = {Oct},
  publisher = {American Physical Society},
  doi = {10.1103/PhysRevA.92.042122},
  url = {https://link.aps.org/doi/10.1103/PhysRevA.92.042122}
}

@article{Rosiek2023,
      title={Quadrature squeezing enhances Wigner negativity in a mechanical Duffing oscillator}, 
      author={Christian A. Rosiek and Massimiliano Rossi and Albert Schliesser and Anders S. Sørensen},
      year={2023},
      journal={arXiv: 2312.12986},
      archivePrefix={arXiv},
      primaryClass={quant-ph},
      url = {https://arxiv.org/abs/2312.12986}
}

\end{document}